\pdfoutput=1
\documentclass[
  aps,
  prx,
  twocolumn,
  superscriptaddress,
  10pt,
  floatfix
]{revtex4-2}

\usepackage[T1]{fontenc}
\usepackage[utf8]{inputenc}
\usepackage{lmodern}
\usepackage[caption=false]{subfig}
\usepackage[normalem]{ulem}
\usepackage{amsmath,amssymb,amsfonts}
\usepackage{bm}
\usepackage{physics}
\usepackage{braket}
\usepackage{placeins}
\usepackage{graphicx}
\graphicspath{{figures/}}
\usepackage{booktabs}
\usepackage{multirow}

\usepackage[dvipsnames]{xcolor}
\usepackage[
  colorlinks=true,
  linkcolor=MidnightBlue,
  citecolor=magenta,
  urlcolor=blue
]{hyperref}

\usepackage{orcidlink}
\usepackage{algorithm}
\usepackage{algpseudocode}

\usepackage{etoolbox}

\makeatletter
\AtBeginDocument{%
  \patchcmd{\@makecaption}
    {\@ifdim{\wd\@tempboxa >\hsize}}
    {\@firstoftwo}
    {}
    {\PackageWarning{mydocument}{Could not patch caption formatting}}%
}
\makeatother

\begin{document}

% \title{Variational Real-Time Dynamics on Reduced Operator Manifold - Application to Quantum Selected Configuration Interaction}
\title{Variational Real-Time Dynamics on Reduced Operator Manifolds}
\author{Aeishah Ameera Anuar}
\email{aeishah.ameera@iqm.tech}
\affiliation{IQM Quantum Computers, Georg-Brauchle-Ring 23-25, 80992 Munich, Germany}
\affiliation{Technical University of Munich, CIT, Department of Computer Science, Boltzmannstra{\ss}e 3, 85748 Garching, Germany}
\author{PV Sriluckshmy}
\affiliation{IQM Quantum Computers, Georg-Brauchle-Ring 23-25, 80992 Munich, Germany}
\author{Riccardo Rossi}
\affiliation{Institute of Physics,École Polytechnique Fédérale de Lausanne (EPFL), CH-1015 Lausanne, Switzerland}
\affiliation{CNRS, Laboratoire de Physique Théorique de la Matière Condensée, Sorbonne Université, 75005 Paris, France}
\author{Fedor \v{S}imkovic IV}
\affiliation{IQM Quantum Computers, Georg-Brauchle-Ring 23-25, 80992 Munich, Germany}

\date{\today}

\begin{abstract}
Accurate real-time simulation of correlated quantum systems remains challenging for both classical methods and near-term quantum hardware. We introduce operator-projected variational quantum real-time evolution (OVQRTE), which updates a parameterized circuit by enforcing the Ehrenfest equations for a selected set of observables. OVQRTE requires only expectation-value measurements, while the choice of operator set enables a systematic trade-off between accuracy and measurement cost, substantially reducing quantum-resource requirements relative to existing variational real-time-evolution algorithms. After implementing OVQRTE dynamics of Heisenberg model on a simulator, we benchmark the algorithm for the Anderson impurity models on the IQM Emerald superconducting processor using up to 24 qubits. We further use OVQRTE to sample computational-basis states for quantum-selected configuration interaction (QSCI), enabling the calculation of the ground-state energy and density of states within a self-consistent ghost-Gutzwiller Ansatz (gGut) embedding loop. Our results establish OVQRTE as a promising approach for investigating correlated condensed-matter systems on near-term quantum hardware.
\end{abstract}

\maketitle

\section{Introduction}

Simulating the real-time dynamics of correlated quantum many-body systems is a central task across condensed matter, quantum chemistry, and materials science, and a paradigmatic application of quantum computing~\cite{lloyd1996}. Many scalable classical approaches an explicit representation of the exponentially large many-body wavefunction by restricting the dynamics to compressed variational representations. Tensor-network methods, for example, approximate the evolution within a structured low-entanglement manifold, including through the time-dependent variational principle~\cite{haegeman2011,paeckel2019}. Neural quantum states (NQS) provide a complementary representation in which the many-body wavefunction is encoded by a neural network whose parameters can be propagated using time-dependent variational Monte Carlo or related projection schemes~\cite{Carleo_2017,schmitt2020,sini2026}. These methods replace the exponential Hilbert-space representation by different numerical bottlenecks: tensor-network simulations become increasingly costly as entanglement grows~\cite{paeckel2019}, whereas NQS dynamics relies on stochastic sampling and nonlinear variational optimization, whose cost and stability can become limiting as the complexity of the evolving state increases.

On quantum hardware, real-time evolution can be implemented by approximating the propagator. Product-formula methods such as Trotterization are straightforward, but require circuit depths that grow with the simulated time and target accuracy~\cite{lloyd1996,childs2021trotter}. Stochastic alternatives such as qDRIFT replace deterministic product formulas by randomized sequences of Hamiltonian terms sampled according to their strengths and avoid an explicit dependence on the number of Hamiltonian terms, at the cost of averaging over an ensemble of circuit realizations~\cite{campbell2019}. This trade-off is most favorable for Hamiltonians with strongly nonuniform coefficients and can be less advantageous for local condensed-matter Hamiltonians with comparable couplings and additional structure that can be exploited by deterministic product formulas~\cite{paganelli2026randomization,chen2021concentration,childs2019lattice}. Spectral methods based on quantum signal processing, qubitization, and quantum singular-value transformation achieve asymptotically favorable query complexity~\cite{low2017qsp,low2019,gilyen2019}. Their block-encoding and coherence requirements remain challenging for near-term hardware, although recent stochastic and randomized approaches indicate that part of this coherent circuit cost can instead be traded for sampling and repeated measurements~\cite{martyn2025stochastic,wang2025randomizedqsvt,qcpp2026}.

Variational quantum real-time evolution offers a different compromise by restricting the dynamics to a parameterized circuit ansatz~\cite{li2017,yuan2019}. In standard McLachlan-based variational real-time evolution, the Schrödinger equation is projected onto the tangent space of the ansatz, yielding a linear system involving the quantum geometric tensor~\cite{yuan2019}. Repeated estimation and solution of this system can incur substantial measurement overhead and can be sensitive to noise and ill-conditioning. Alternative formulations can avoid explicit construction of the quantum geometric tensor: projected variational quantum dynamics (p-VQD) determines successive states through an overlap-based projection, while dual variational time evolution reformulates the update through a fidelity-based optimization~\cite{barison2021,gacon2024}. These approaches reduce the geometric-tensor bottleneck, but replace it with different requirements associated with overlap estimation and classical optimization.

An alternative approach consists in formulating the variational problem directly in terms of selected observables. Reduced-density-matrix (RDM) methods provide a natural classical example of this approach. For two-body Hamiltonians, the two-particle RDM has polynomial size and contains sufficient information to evaluate the energy and all one- and two-body observables~\cite{mazziotti2006,mazziotti2012,lackner2015}. Its exact dynamics, however, is not self-contained: the equation of motion of the two-particle RDM depends on the three-particle RDM, leading to a hierarchy that must be approximately closed in practical classical propagation. Moreover, an approximately propagated RDM need not correspond to a physical (N)-particle state, requiring control or enforcement of (N)-representability~\cite{lackner2015,mazziotti2012}.
A parameterized quantum circuit enables a different realization of this observable-level perspective. At the level of the underlying noiseless variational state, it represents a physical many-body state by construction, while higher-body expectation values entering the equations of motion can be measured directly rather than reconstructed through a closure functional. The resulting ideal-state evolution therefore requires neither an explicit closure of the RDM hierarchy nor a separate representability correction. Experimentally estimated RDMs may nevertheless violate representability constraints because of finite sampling and hardware noise, so representability-based post-processing can remain useful in practice~\cite{rubin2018}.

The utility of approximate real-time evolution extends beyond the direct calculation of dynamical observables. Sets of time-evolved reference states naturally generate quantum Krylov-like subspaces in which low-lying eigenstates with nonzero reference-state overlap can be approximated by subspace diagonalization, motivating quantum-classical approaches that use dynamics to construct compact spaces for ground- and excited state calculations~\cite{parrish2019qfd,klymko2022,cortes2022}. Quantum-selected configuration interaction (QSCI) instead constructs a compact configuration space by sampling important many-body configurations from a quantum-prepared state and subsequently diagonalizing the Hamiltonian in the selected sub-space~\cite{qsci}. Time-evolution-based variants, including TE-QSCI and Hamiltonian-simulation-based QSCI, generate these sampled configurations from states evolved under the target Hamiltonian~\cite{mikkelsen2025,sugisaki2025}. Related work uses stochastic Hamiltonian time evolution and time-evolved population statistics to guide the construction and expansion of compact selected-CI spaces~\cite{weaving2026timeevolvedsci}. SqDRIFT is more naturally viewed as a randomized extension of sample-based Krylov quantum diagonalization (SKQD): SKQD samples configurations from quantum Krylov states, whereas SqDRIFT generates these states using qDRIFT-randomized approximations to the Hamiltonian propagator, reducing the circuit-depth requirements of the time-evolution step~\cite{yu2025skqd,piccinelli2025}. Collectively, these approaches illustrate that real-time evolution can remain useful even when the complete dynamical trajectory is not reproduced to high accuracy, provided that the evolved states generate an informative subspace for the downstream calculation~\cite{sugisaki2025,piccinelli2025}.

In this work, we develop Operator-Projected Variational Quantum Real-Time Evolution (OVQRTE), the real-time counterpart of the operator-projected variational framework introduced for imaginary-time evolution in OVQITE~\cite{ovqite}. Rather than projecting the complete state evolution onto the ansatz tangent space or determining successive states through a global state-overlap objective, OVQRTE determines the parameter updates by matching the exact Ehrenfest equations for a chosen set of observables. This set may consist, for example, of local Pauli strings or the operators defining the one- and two-particle RDMs, and its size provides a systematic knob for trading accuracy against circuit depth and measurement cost. The update requires only expectation values and avoids estimation of the quantum geometric tensor. We first derive the OVQRTE equations, analyze their quantum and classical costs, and introduce a procedure for reducing the tracked operator set. We test the implementation on the one-dimensional Heisenberg model, where its scaling and noise resilience are particularly favorable, and on a fermionic impurity model. Finally, motivated by the use of real-time dynamics in quantum subspace methods, we demonstrate OVQRTE on quantum hardware as a generator of states for subspace-based ground-state estimation.

The paper is structured as follows: We introduce the relevant theory in Sec.~\ref{sec:theory}, and showcase simulated results for the Heisenberg model Sec.~\ref{sec:heisenberg} and quantum hardware of OVQRTE for the Anderson Impurity model and we discuss the QSCI calculation for the ground-state energy and density of states Sec.~\ref{sec:aim} before providing concluding remarks in Sec.~\ref{sec:conclusions}.

\section{Theory}\label{sec:theory}

% ============================================================
%  OVQRTE — theory section, rewritten (concise)
%  Footing: projection + ansatz expressivity (NOT closure).
%  Trajectory accuracy controlled by invariance of S under [H,.].
%  Requires: amsmath, amssymb, amsthm, bm
%  Preamble needs: \newtheorem{proposition}{Proposition}
% ============================================================

\subsection*{Variational real-time dynamics on reduced operator manifolds}

The state of the system is described by a density matrix $\rho(t)$, which
evolves according to the Liouville-von Neumann equation
\begin{equation}
\label{eq:density}
    \dot{\rho}(t)=-i[\hat H,\rho(t)].
\end{equation}
Equivalently, $\rho(t)$ can be characterized by its expectation values with
respect to a complete Hermitian operator basis $\{\hat O_\alpha\}$,
\begin{equation}
    x_\alpha(t)
    =
    \langle \hat O_\alpha\rangle_t
    =
    \operatorname{Tr}\!\left[\rho(t)\hat O_\alpha\right].
    \label{eq:operator-coefficients}
\end{equation}
The complete set of expectation values $\{x_\alpha(t)\}$ therefore uniquely
determines the density matrix. For an $n$-qubit system, however, a complete
operator basis contains $\mathcal{O}(4^n)$ elements and thus grows
exponentially with the system size.

The exact dynamics of each coefficient follows directly from \ref{eq:density} and is given by the Ehrenfest equation
\begin{equation}
    \dot{x}_\alpha
    =
    \frac{d}{dt}\langle\hat O_\alpha\rangle
    =
    i\langle[\hat H,\hat O_\alpha]\rangle.
    \label{eq:ehrenfest}
\end{equation}
Rather than retaining the complete operator basis, we select a
polynomially sized subset of Hermitian operators,
\begin{equation}
    \mathcal{S}
    =
    \{\hat O_\alpha\}_{\alpha=1}^{|\mathcal{S}|},
    \qquad
    |\mathcal{S}|=\mathrm{poly}(n),
\end{equation}
and track only the corresponding expectation values
 $x_\alpha(t)=\langle\hat O_\alpha\rangle_{\rho(t)}$.  The operators in $\mathcal{S}$ are meant to furnish a basis for a reduced representation of the evolving state. Concretely, we select $\mathcal{S}$ by operator-weight or locality: since the commutator in Eq.~\eqref{eq:ehrenfest} raises the weight of an operator by at most that of $\hat H$ itself, the set of operators up to a fixed weight stays approximately self-contained under a $k$-local or few-body Hamiltonian, and it is this set, the 1- and 2-body RDM operators, or Pauli strings of bounded weight, that we track. Because such sets grow only as $O(n^2)$-$O(n^4)$ at fixed operator weight, or $O(n^\ell)$ at fixed Pauli weight $\ell$, rather than the full $4^n$ operators needed to specify $\rho(t)$ exactly, $|\mathcal{S}|$ scales polynomially with system size by construction. 
% \textcolor{gray}{This mirrors the operator-space restriction of our imaginary-time formulation, OVQITE~\cite{ovqite}, though there an additional safeguard exists: any admissible $\mathcal{S}$ still drives the energy monotonically downward, a scalar witness of progress that real-time evolution lacks.}

% ---- variational manifold (general, then specialized) ----
The construction so far is state-agnostic: it applies to any family of parameterized states
$\rho(\bm\theta)$, $\bm\theta\in\mathbb{R}^p$, that induces the observable map
$x_\alpha(\bm\theta)=\langle\hat O_\alpha\rangle_{\rho(\bm\theta)}$, whether
$\rho(\bm\theta)$ is a classical variational ansatz (a mean-field or
tensor-network parameterization) or the output of a quantum circuit. The key modelling choice is where the dynamics is imposed: rather than projecting the full state evolution, we impose the Ehrenfest equation~\eqref{eq:ehrenfest} only on the operators in $\mathcal{S}$, and use this restricted requirement to fix the parameter trajectory. In this work, we realize $\rho(\bm\theta)$ with a parameterized
quantum circuit,
$\rho(\bm\theta)=|\psi(\bm\theta)\rangle\langle\psi(\bm\theta)|$, whose
advantage is that expectation values of operators beyond the reach of classical
ans\"atze remain efficiently accessible. Since the observables change solely
through parameter motion,
\begin{equation}
    \dot x_\alpha^{\rm var}=\sum_{j=1}^p \dot\theta_j\,g_{\alpha j},
    \qquad
    g_{\alpha j}=2\,\Re\langle\psi|\hat O_\alpha|\partial_j\psi\rangle\in\mathbb{R},
\end{equation}
where the superscript ``var'' marks the rate of change induced by the
variational trajectory, to be distinguished from the exact rate in
Eq.~\eqref{eq:ehrenfest}, we define the residual
$R_\alpha(\dot{\bm\theta})=\dot x_\alpha^{\rm var}-i\langle[\hat H,\hat O_\alpha]\rangle$
and determine the parameter update $\dot{\bm\theta}$ by a least-squares fit,
\begin{equation}
    \dot{\bm\theta}=\arg\min_{\mathbf v\in\mathbb{R}^p}
    \sum_{\alpha=1}^{|\mathcal{S}|} \bigl|R_\alpha(\mathbf v)\bigr|^2 .
    \label{eq:leastsquares}
\end{equation}
Because $[\hat H,\hat O_\alpha]$ is anti-Hermitian,
$i\langle[\hat H,\hat O_\alpha]\rangle$ is real; the driving terms are
expectation values measured on $\rho(\bm\theta)$, computed exactly up to
sampling and hardware noise, and need not lie in $\mathrm{span}(\mathcal{S})$. The minimization
yields the linear system $A\,\dot{\bm\theta}=C$ with
\begin{align}
    A_{jk}&=\sum_{\alpha=1}^{|\mathcal{S}|}g_{\alpha j}g_{\alpha k},\\
    C_{j}&=\sum_{\alpha=1}^{|\mathcal{S}|}g_{\alpha j}\,i\langle[\hat H,\hat O_\alpha]\rangle,
\end{align}
which we solve for $\dot{\bm\theta}(t)$ at each step and advance the
parameters via an explicit Euler update,
\begin{equation}
    \bm\theta(t+\delta t)=\bm\theta(t)+\delta t\,\dot{\bm\theta}(t).
    \label{eq:euler}
\end{equation}
In practice, we solve the linear system for $\dot{\bm\theta}(t)$ using a
regularized least-squares solver, which truncates small
singular values of $A$ to guard against ill-conditioning and divergence,
rather than inverting $A$ directly.
Here, $A=G^\top G$ is the Gram matrix of the observable Jacobian
$G_{\alpha j}=(g_{\alpha j})$, not the quantum geometric tensor that governs standard variational time
evolution\cite{yuan-theory}. This is the real-time analogue of
the imaginary-time OVQITE construction of Ref.~\cite{ovqite}, which we
name Operator-Projected Variational Quantum Real-Time Evolution (OVQRTE).
%
% ---- the two approximations ----
% \paragraph{Two independent approximations.}
The method rests on two logically distinct approximations, controlled by two independent knobs:

\emph{(i) Information restriction.} We demand fidelity of the dynamics only for
the tracked expectation values $\{x_\alpha\}_{\alpha=1}^{|\mathcal{S}|}$ associated with the chosen set
$\mathcal{S}$. Formally this is a \emph{projection}:
for an orthonormal $\mathcal{S}$,
\begin{equation}
    \sum_{\alpha=1}^{|\mathcal{S}|} |R_\alpha(\dot{\bm\theta})|^2
    = \Bigl\|\mathcal{P}_{\mathcal S}\Bigl(\dot\rho(\bm\theta)+i[\hat H,\rho(\bm\theta)]\Bigr)\Bigr\|_{\rm HS}^2 ,
\end{equation}
where $\dot\rho(\bm\theta)=\sum_j\dot\theta_j\,\partial_{\theta_j}\rho(\bm\theta)$ and
$\mathcal{P}_{\mathcal S}$ is the orthogonal projector onto $\mathrm{span}(\mathcal S)$
under the (normalized) Hilbert-Schmidt inner product. So Eq.~\eqref{eq:leastsquares}
minimizes the component of the exact McLachlan residual visible to $\mathcal{S}$ and
discards its orthogonal complement.

\emph{(ii) Variational realization.} 
The induced expectation values define the map $\mathbb{R}^p\to\mathbb{R}^{|\mathcal{S}|}$
whose image
\begin{equation}
    \mathcal{M}_{\mathrm{op}}
    = \bigl\{ \bigl(x_1(\bm\theta),\dots,x_{|\mathcal{S}|}(\bm\theta)\bigr):\bm\theta\in\mathbb{R}^p\bigr\}
    \subset\mathbb{R}^{|\mathcal{S}|}
\end{equation}
we call the \emph{variational operator manifold}.
The tracked expectation values do not evolve freely:
their entire trajectory is realized through the parameter path $\bm\theta(t)$ on
$\mathcal{M}_{\rm op}$, so only trajectories reachable within this manifold of
dimension at most $p$ can be represented, same as in any variational method. Crucially,
 the exact Ehrenfest rates
$i\langle[\hat H,\hat O_\alpha]\rangle$ entering $R_\alpha$ are evaluated directly on physical state $\rho(\theta)$ up to sampling and hardware noise. 
% This remains possible even when the commutator lies outside of the selected operator set $\mathcal{S}$ and no reconstruction or projection of the commutator on $\mathcal{S}$ is required.

% never reconstructed or
% approximated, even though the operator $[\hat H,\hat O_\alpha]$ generally lies outside
% $\mathcal{S}$ as they are evaluated directly as expectation values on the physical state
% $\rho(\bm\theta)$, exact up to sampling and hardware noise. 

% A quantum realization moreover cannot violate
% $N$-representability, since every $\rho(\bm\theta)$ produced by a quantum circuit is a
% valid physical state by construction, in contrast with classical
% reduced-density-matrix propagation, whose trajectory can leave the physical set.

Both approximations can be removed independently: enlarging $\mathcal{S}$ toward a
complete operator basis eliminates approximation (i), recovering full McLachlan evolution;
enlarging the ansatz toward universality eliminates approximation (ii).

At a fixed time $t$, solving Eq.~\eqref{eq:leastsquares} makes every tracked
derivative exact for the current state,
\begin{equation}
    \dot x_\alpha^{\rm var} = i\langle[\hat H,\hat O_\alpha]\rangle_{\rho(\bm\theta(t))}
    \qquad\text{for all }\alpha,
\end{equation}
regardless of whether $[\hat H,\hat O_\alpha]\in\mathrm{span}(\mathcal{S})$. The trajectory
accuracy is a different question, because the right-hand side depends on the entire
state $\rho(\bm\theta(t))$, and only its $\mathcal{S}$-component is constrained by the
fit and the untracked component evolves under its own unimposed Ehrenfest equation and can
drift, feeding back into the tracked derivatives at later times. Let's consider here the idealized limit where $\mathcal{S}$ is invariant under the
adjoint action of $\hat H$,
\begin{equation}
    i[\hat H,\hat O_\alpha]\in\mathrm{span}(\mathcal{S})
    \quad\text{for all }\alpha,
    \label{eq:invariance}
\end{equation}
so that $i[\hat H,\hat O_\alpha]=\sum_\beta c_{\alpha\beta}\hat O_\beta$ with real
coefficients $c_{\alpha\beta}$.  The Ehrenfest driving term then
closes within the tracked set,
\begin{equation}
    i\langle[\hat H,\hat O_\alpha]\rangle
    =
    \sum_\beta c_{\alpha\beta}x_\beta,
\end{equation}
and is therefore determined entirely by the tracked expectation values,
independent of the untracked sector.
% We stress
% that Eq.~\eqref{eq:invariance} is not a design criterion as we do not inspect
% commutators in the scaling of $\mathcal{S}$ but it is an idealized limit that explains why the method succeeds when it does.
Since repeatedly commuting a local operator with $\hat H$ generically spreads its
support until it spans an exponentially large subspace, the smallest invariant
$\mathcal{S}$ containing a local observable is exponentially large, so exact
invariance is unattainable at polynomial cost, and what matters in practice is
how nearly it holds over the times of interest.

\subsection*{Quantum resource requirements}

Each OVQRTE step solves the linear system $A\,\dot{\bm\theta}=C$, and its cost
is set by measuring the two ingredients, $A=G^\top G$ and $C$, on the current
state. The matrix $A$ is built from the observable Jacobian
$g_{\alpha j}=\partial_{\theta_j}\langle\hat O_\alpha\rangle$, which we evaluate
by finite differences: shifting each parameter forward and backward gives $2p$
circuit settings, and at each setting all $|\mathcal{S}|$ tracked operators are measured
\emph{simultaneously} by sorting them into $\operatorname{C}_{\mathrm{pool}}$ qubit-wise
commuting (QWC) groups, leading to $2p\operatorname{C}_{\mathrm{pool}}$ circuits overall. The vector
$C$ follows from $C_j=\sum_\alpha g_{\alpha j}\,i\langle[\hat H,\hat O_\alpha]\rangle$,
whose commutator expectation values are measured once at the current parameters
in $\operatorname{C}_{\mathrm{comm}}$ QWC groups. The total circuit count per iteration is
therefore
\begin{equation}
    \operatorname{C}_{\mathrm{total}} 
    = 2p\operatorname{C}_{\mathrm{pool}} +  \operatorname{C}_{\mathrm{comm}}.
    \label{eq:cost-fd}
\end{equation}
% \begin{equation}
%     C_{\mathrm{total}}
%     = \underbrace{2p\,C_{\mathrm{pool}}}_{\text{metric }A}
%     + \underbrace{C_{\mathrm{comm}}}_{\text{vector }C}.
%     \label{eq:cost-fd}
% \end{equation}

The $2p$ prefactor can be removed by estimating the Jacobian with the
simultaneous perturbation stochastic approximation (SPSA)\cite{Spall1998,gacon-qgt-free}, which perturbs all
$p$ parameters along a single random direction per sample, requiring only two
evaluations per sample. With $N_s$ samples, the total cost becomes
$2N_s \operatorname{C}_{\mathrm{pool}}$,
\begin{equation}
    \operatorname{C}_{\mathrm{total}}^{\mathrm{(SPSA)}}
    = 2N_s\operatorname{C}_{\mathrm{pool}} + \operatorname{C}_{\mathrm{comm}} ,
    \label{eq:cost-spsa}
\end{equation}
where $N_s$ is a user-chosen constant (typically $N_s\!\sim\!10$--$20$) that
need not grow with $p$ or $n$, though it may be increased with system size if
higher accuracy is required. The finite-difference and SPSA estimators are
detailed in Appendix~\ref{app:implementation}.

The essential advantage of OVQRTE is the \emph{absence} of the quantum geometric tensor. In
the related variational real-time evolution based on Mclachlans principle (VQRTE) method~\cite{li2017,yuan2019,barison2021}, the parameter update is governed by the
quantum geometric tensor (QGT), whose $O(p^2)$ entries require the execution of state-overlap (Hadamard- or
SWAP-test) at a cost that
grows quadratically in the number of parameters (see Appendix \ref{app:implementation}). OVQRTE replaces the evaluation of QGT with
with $A=G^\top G$, assembled entirely from expectation values of the tracked
observables and their parameter derivatives where no overlap circuits are needed and the
per-setting measurements are shared across all operators through QWC grouping,
and the circuit count scales \emph{linearly} in $p$ (Eq.~\ref{eq:cost-fd}), or
independently of $p$ up to the constant $N_s$ under SPSA
(Eq.~\ref{eq:cost-spsa}). The grouping counts $\operatorname{C}_{\mathrm{pool}}$ and
$\operatorname{C}_{\mathrm{comm}}$ are governed by the locality of $\mathcal{S}$ and of the
commutators $[\hat H,\hat O_\alpha]$, and remain modest for the local operator
sets used here. Equivalently, the same few-body expectation values may instead
be obtained simultaneously from classical shadows, with sample complexity
$O(3^{\ell'}\log |\mathcal{S}|/\epsilon^2)$ at locality $\ell'=k+\ell-1$,  and precision
$\epsilon$~\cite{huang-shadows}, independent of $n$ assuming that the Hamiltonian is $k$-local and the operators in $\mathcal{S}$ are at most $\ell$-local.

% \noindent The asymptotic comparison for QEB pools becomes:
% \begin{center}
% \begin{tabular}{l c c}
% \toprule
% Component & FD  & SPSA  \\
% \midrule
% Gradient matrix & $\mathcal{O}(p) = \mathcal{O}(n)$ & $\mathcal{O}(N_s) = \mathcal{O}(1)$ \\
% Evolution vector & $\mathcal{O}(n^2)$ & $\mathcal{O}(n^2)$ \\
% \midrule
% \textbf{Total} & $\mathcal{O}(n^2)$ & $\mathcal{O}(n^2)$ \\
% \bottomrule
% \end{tabular}
% \end{center}

% When considering quantum circuit

\subsection*{QEB-inspired operator pool}

In fermion-to-qubit mappings such as the Jordan-Wigner transformation, excitation operators generally contain nonlocal parity strings composed of Pauli-$Z$ operators. These strings increase the support of the mapped operators, leading to deeper circuits and making the evaluation of expectation values more susceptible to noise. Motivated by qubit-excitation-based (QEB) constructions \cite{qeb}, we can therefore replace $Z$ Paulis within Pauli strings with identity operators $I$, leading to a significant reduction in resource requirements. This reduction can be viewed as the simplification of Jordan-Wigner excitation operators in which the intermediate non-locality is entirely due to fermionic parity strings. More precisely, for one-body fermionic terms mapped to spin operators, one replaces Pauli strings as
\begin{equation}
\label{eq:QEB}
    P_{ij}^{\mu\nu}
    =
    \sigma_i^\mu
    \left(
        \prod_{k=i+1}^{j-1} Z_k
    \right)
    \sigma_j^\nu  \;\longmapsto\;
    \sigma_i^\mu \sigma_j^\nu,
\end{equation}
with endpoint operators \(\sigma_i^\mu,\sigma_j^\nu  \in \{X,Y\}\). 
The resulting reduced pool should be regarded as a qubit-excitation-inspired approximation rather than an exact preservation of the original fermionic algebra. Its purpose is to construct a hardware-friendlier operator set with lower locality overhead, leading to shorter Pauli strings for measurement. Consequently, this reduction is expected to yield more favorable scaling with system size and improved robustness against noise in expectation value estimation.

\subsection*{Ground-state extraction via QSCI}

Finally, the approximate real-time evolution generated by OVQRTE can be repurposed for spectral estimation rather than used only to evaluate dynamical observables. Given an initial state $|\Psi_0\rangle$, a collection of states along the real-time trajectory,
$
|\Psi(t_j)\rangle \simeq e^{-i\hat{\mathcal H}t_j}|\Psi_0\rangle,
$
spans a real-time Krylov-like subspace
$
\mathcal{K}_{T}=\mathrm{span}\{|\Psi(t_0)\rangle,|\Psi(t_1)\rangle,\ldots\}
$~\cite{parrish2019qfd,klymko2022,cortes2022}.
For equally spaced times $t_j=j\Delta t$, this can be viewed as a unitary Krylov space generated by repeated application of
$\hat U(\Delta t)=e^{-i\hat{\mathcal H}\Delta t}$; correspondingly, expanding $\hat U(\Delta t)$ in powers of $\hat{\mathcal H}$ connects the real-time construction to the conventional polynomial Krylov space generated by
$\{|\Psi_0\rangle,\hat{\mathcal H}|\Psi_0\rangle,\hat{\mathcal H}^2|\Psi_0\rangle,\ldots\}$.
Consequently, provided that $|\Psi_0\rangle$ has nonzero overlap with the relevant low-energy eigenstates, the dynamical trajectory contains information that can be used to approximate those eigenstates through subspace diagonalization~\cite{parrish2019qfd,klymko2022,cortes2022}. In the present work, rather than directly measuring the overlap and Hamiltonian matrices between the generally nonorthogonal time-evolved states, we use the OVQRTE trajectory to generate a compact \emph{configuration} subspace through quantum-selected configuration interaction (QSCI)~\cite{qsci}. This strategy is closely related to time-evolution- and Hamiltonian-simulation-based variants of QSCI~\cite{mikkelsen2025,sugisaki2025}, as well as sample-based Krylov quantum diagonalization (SKQD), which combines sampling from quantum Krylov states with classical subspace diagonalization~\cite{yu2025skqd}.

% Specifically, computational-basis measurements of the OVQRTE states at the sampled evolution times produce a set of configurations
% $\mathcal{X}=\{|\mathbf{x}_k\rangle\}$. We define
% \begin{equation}
%     \mathcal{C} = \mathrm{span}\{|\mathbf{x}_k\rangle:\mathbf{x}_k\in\mathcal{X}\},
%     \qquad
%     \hat P_{\mathcal C}
%     =
%     \sum_{k}
%     |\mathbf{x}_k\rangle\langle\mathbf{x}_k|.
% \end{equation}

Specifically, computational-basis measurements of the OVQRTE state at
each sampled evolution time $t_j$ produce a per-iteration configuration
set $\mathcal{X}_j=\{|\mathbf{x}_k\rangle\}$; the cumulative set used for
the subspace at iteration $T$ is their union over all iterations reached
so far,
\begin{equation}
    \mathcal{X}_T = \bigcup_{j=0}^{T}\mathcal{X}_j,
\end{equation}
so that $\mathcal{X}_0$ is the first iteration's samples on their own and
each subsequent $\mathcal{X}_T$ adds to, rather than replaces, those
already collected. We define
\begin{equation}
    \mathcal{C} = \mathrm{span}\{|\mathbf{x}_k\rangle:\mathbf{x}_k\in\mathcal{X}_T\},
    \qquad
    \hat P_{\mathcal C}
    =
    \sum_{k}
    |\mathbf{x}_k\rangle\langle\mathbf{x}_k|.
\end{equation}
The ground state is then approximated variationally by diagonalizing the Hamiltonian projected into this sampled subspace,
\begin{equation}
    \hat{P}_{\mathcal{C}} \hat{\mathcal{H}} \hat{P}_{\mathcal{C}}
    |\Psi^{\mathcal{C}}_{\mathrm{GS}}\rangle
    =
    E_{\mathrm{GS}}^{\mathcal{C}}
    |\Psi^{\mathcal{C}}_{\mathrm{GS}}\rangle,
    \label{eq:qsci_diag}
\end{equation}
yielding the QSCI energy estimate $E_{\mathrm{GS}}^{\mathcal{C}}$ and its corresponding approximate ground state wavefunction
$|\Psi^{\mathcal{C}}_{\mathrm{GS}}\rangle$~\cite{qsci}.
Thus, accurate reproduction of the complete real-time trajectory is not required for this downstream use of OVQRTE: it is sufficient that the evolved states place appreciable measurement weight on configurations that collectively span an informative low-energy subspace.

On noisy hardware, however, errors in state preparation and measurement can redistribute this sampled weight and, in particular, produce bitstrings outside symmetry sectors that are exactly preserved by the target Hamiltonian. Such symmetry violations provide information that can be exploited during classical post-processing, in the same general spirit as symmetry-based quantum error mitigation~\cite{bonetmonroig2018symmetry}. 

For many fermionic Hamiltonians from condensed matter physics and quantum chemistry, the particle numbers of the two fermionic species are separately conserved, so ideal samples have fixed Hamming weights $(\mathcal{N}_\alpha,\mathcal{N}_\beta)$. Rather than simply discarding samples that violate these constraints, we apply the self-consistent configuration recovery (S-CORE) procedure of Ref.~\onlinecite{recovery}. S-CORE probabilistically maps configurations with incorrect particle numbers back into the target $(\mathcal{N}_\alpha,\mathcal{N}_\beta)$ sector, using the average spin-orbital occupations $\bar n_{p\sigma}$ obtained from the current CI solution to preferentially select the bit flips used in the recovery. The recovered configurations are subsequently rediagonalized, the occupations are updated, and the recovery can be iterated self-consistently~\cite{recovery}. In this sense, S-CORE acts as a symmetry-informed error-mitigation layer on the sampling stage while retaining information from noisy outcomes that would otherwise be discarded.

In this paper, we additionally exploit the particle-hole (PH) symmetry of the Anderson impurity model (AIM), which we present further in the text, at half filling by explicitly closing the recovered configuration set under the PH transformation. Denoting the recovered set by $\mathcal{X}_{\mathrm R}$ and the particle-hole partner of $|\mathbf{x}_k\rangle$ by $|\tilde{\mathbf{x}_k}\rangle$, we construct
\begin{equation}
    \mathcal{X}_{p}
    =
    \mathrm{unique}
    \left[
        \mathcal{X}_{\mathrm R}
        \cup
        \widetilde{\mathcal{X}}_{\mathrm R}
    \right],
    \qquad
    \mathcal{C}_{p}
    =
    \mathrm{span}\{|\mathbf{x}_k\rangle:\mathbf{x}_k\in\mathcal{X}_{p}\},
\end{equation}
with projector
\begin{equation}
    \hat P_{\mathcal C_p}
    =
    \sum_{k}
    |\mathbf{x}_k\rangle\langle\mathbf{x}_k|.
\end{equation}
This PH completion is conceptually distinct from S-CORE: S-CORE mitigates noise-induced violations of the conserved particle-number sector, whereas PH completion deterministically augments the selected space so that it is closed under an exact symmetry of the half-filled Hamiltonian. The resulting S-CORE$+$PH subspace is then used in Eq.~\eqref{eq:qsci_diag} for the final ground-state estimate.

\section{Implementation}

\subsection{The Heisenberg model}
\label{sec:heisenberg}
First, we investigate the real-time evolution of the open boundary 1d Heisenberg model with nearest neighbor interaction $J=1/2$ and transverse field $h=-1$:
\begin{equation}
		\label{heisenberg}
		\hat{H}=J\sum_i (\hat{X}_i \hat{X}_{i+1}+ \hat{Y}_i \hat{Y}_{i+1}+\hat{Z}_i \hat{Z}_{i+1})+h\sum_i\hat{Z}_i.
\end{equation} 
 
\begin{figure}[t]

    % Left column: ansatz + pool scaling vertically stacked

    % Right column: larger VQRTE comparison
        \centering
        \includegraphics[width=1\linewidth]{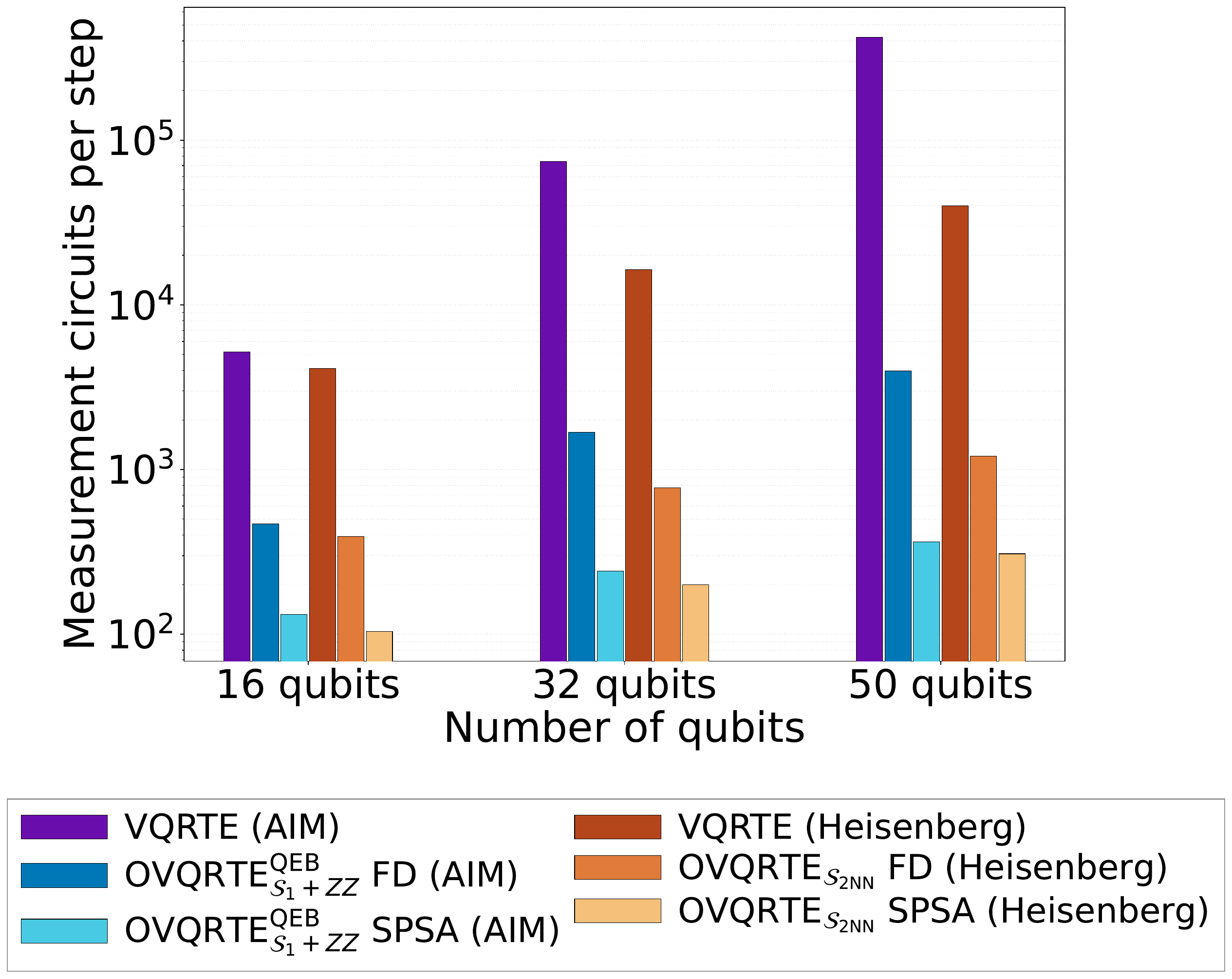}
        
\caption{\textbf{Measurement circuits per time step versus system size}
    ($16$, $32$, $50$ qubits; log scale). VQRTE and OVQRTE using
    finite-difference (FD) and SPSA to compute gradients, for the Heisenberg model, using the HEA ansatz with $L=1$ and $\mathcal{S}_{\mathrm{2NN}}$ operator set, and the Anderson impurity model (AIM), using the LUCJ ansatz with $L=1$ ansatz layers,
    and $\mathcal{S}_{1{+}ZZ}$ operator set.}
    \label{fig:scaling}
\end{figure}

For this model we consider the following operator set,
\begin{equation} \label{eq:operator_set_NN}
\mathcal{S}_{2\text{NN}}\equiv\bigcup_{\alpha, j}\{\hat{P}_{\alpha;j}\}\bigcup_{\alpha,\gamma,\braket{j,k}}\{\hat{P}_{\alpha;j}\hat{P}_{\gamma;k}\},
\end{equation}
where $\hat{P}_{\alpha;j}$ is the $\alpha$-th Pauli operator acting on the qubit at lattice site $j$ and $\braket{j,k}$ means that the sites $j$ and $k$ are nearest neighbors on the lattice.  
As a variational ansatz, we use a circuit with
Pauli-$Y$ and Pauli-$Z$ single qubit rotation layers that alternate
with pairwise $CX$ entangling gates. The circuit is provided in Appendix \ref{fig:hea}. For $n$ qubits, the total number of tunable parameters is $N_\theta=2n(L+1)$ where $L$ is the number of repetition layers. For initialization, we start with the  $|+\rangle^{\otimes n}$ state by setting the first layer of Pauli-$Y$ rotations to $\pi/2$ and the remaining layers to $0$. We use equation \eqref{eq:cost-fd} to plot the scaling of the per-iteration cost of this model with the chosen ansatz and  operator set, and contrast that with the original VQRTE algorithm~\cite{li-benjamin} in Fig.~\ref{fig:scaling}. The result shows that OVQRTE substantially reduces the measurement cost for this model compared to VQRTE, requiring roughly one to two orders of magnitude fewer measurement circuits per time step. The advantage becomes more pronounced with increasing system size, with the SPSA cost achieving close to two order of magnitude reduction for the largest systems considered.

To isolate the effect of operator-set selection on the accuracy of the simulated dynamics, we perform state-vector simulations for small system sizes, which remove statistical errors associated with finite-shot sampling. The resulting dynamics are then benchmarked against those obtained with the original VQRTE formulation, whose implementation is described in detail in Appendix~\ref{app:implementation}. 
 
\begin{figure}[t]

    % Left column: ansatz + pool scaling vertically stacked

    % Right column: larger VQRTE comparison
        \centering
    \includegraphics[width=1\linewidth]{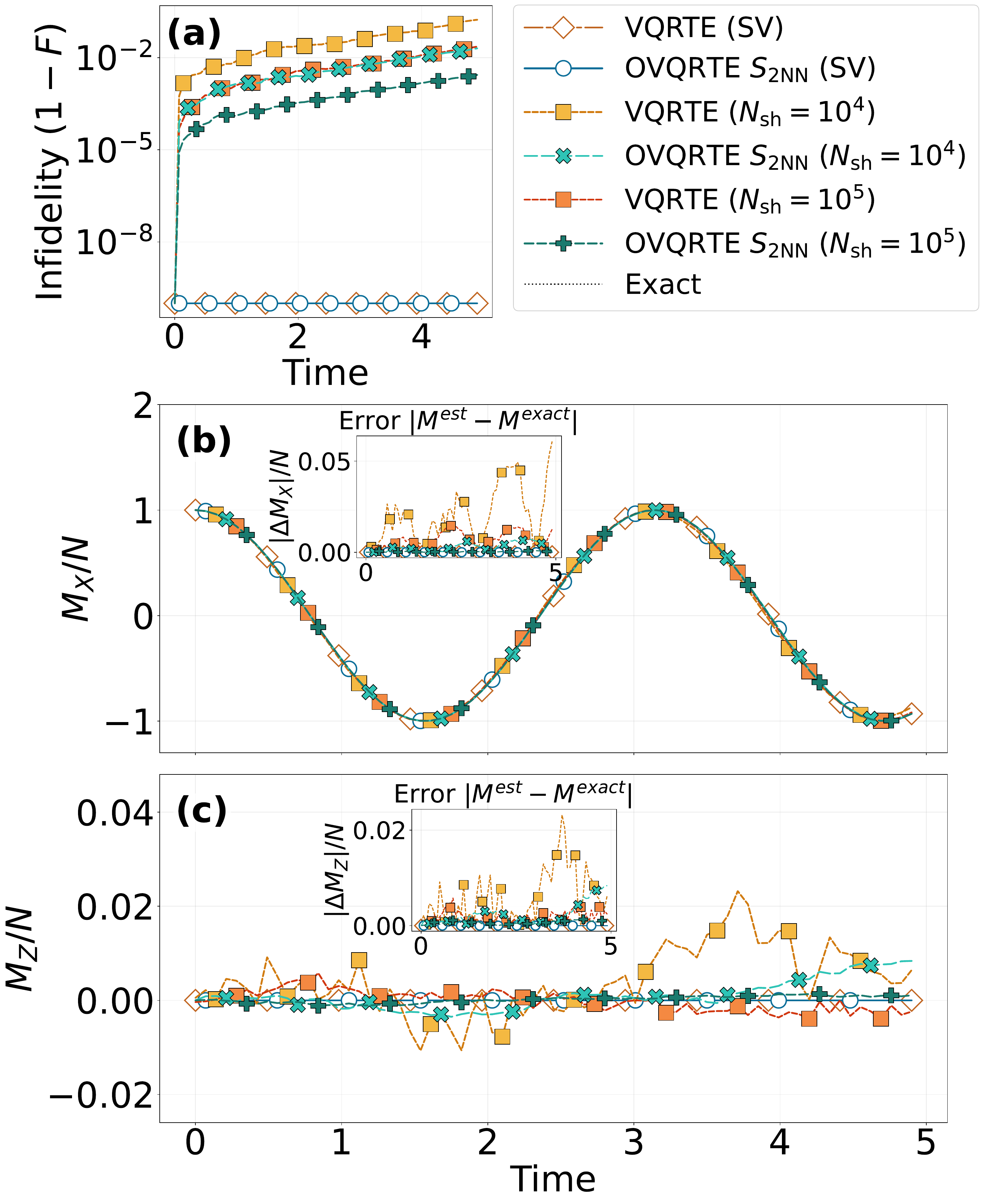}
        
\caption{\textbf{Real-time dynamics of a $12$-qubit Heisenberg chain.}
    % MYCOMMENT: add couplings J, h/J, boundary conditions, initial state.
    OVQRTE projecting onto the two-local nearest-neighbor Pauli set
    $\mathcal{S}_{\mathrm{2NN}}$ versus McLachlan VQRTE, in the statevector
    (SV) limit and at $N_{\text{sh}}=10^4,10^5$. (a) state infidelity $1-F$;
    (b) transverse magnetization $M_X={1/n}\sum_i\langle X_i\rangle$;
    (c) longitudinal magnetization $M_Z={1/n}\sum_i\langle Z_i\rangle$. Exact result dotted.}
    \label{fig:heisenberg}
\end{figure}

We then sample a finite number of measurement shots, $N_{\text{sh}}$, from the ideal expectation values to facilitate realistic statistical measurement noise.
Figure~\ref{fig:heisenberg} (a) shows the state infidelity $1-F$ with respect to exact evolution at different times. In the statevector limit, projecting the dynamics onto $\mathcal{S}_{\mathrm{2NN}}$
reproduces the exact trajectory to numerical precision (with infidelity
$1-F<10^{-9}$) for both OVQRTE and VQRTE. 
With finite
sampling the infidelity grows with time and saturates at a level set by the
number of shots. For OVQRTE we observe a late-time plateau of
$1-F\sim10^{-2}$ for $N_{\text{sh}}=10^4$ and $1-F\sim10^{-3}$ for $N_{\text{sh}}=10^5$
(Fig.~\ref{fig:heisenberg} (a)), which decreases as expected through additional
sampling. VQRTE saturates roughly one order of magnitude higher at each
shot budget, so that OVQRTE at $N_{sh}=10^4$ already reaches the fidelity
VQRTE only attains at $N_{sh}=10^5$. Figure~\ref{fig:heisenberg} (b) and (c), respectively, shows the transverse and longitudinal magnetizations
$M_X={1/n}\sum_i\langle X_i\rangle$ and $M_Z={1/n}\sum_i\langle Z_i\rangle$, normalized by number of qubits. The transverse magnetization $M_X$, which oscillates over the full range
$\pm1$, is tracked closely by both methods at the scale of the main
panel. The inset reveals a sharper distinction, however, with VQRTE evolution
point-wise error growing over time and reaching $|\Delta M_X|\sim0.75$ for
$N_{\text{sh}}=10^4$, while OVQRTE's error remains an order of magnitude smaller
across both shot budgets. The longitudinal magnetization $M_Z$, conserved
and exactly zero for this evolution, provides a more stringent noise
diagnostic with
($|\Delta M_Z|\lesssim0.2$ vs.\ $|\Delta M_X|\lesssim0.75$). This indicates that OVQRTE is more robust to finite-shot sampling noise than VQRTE, reproducing the relevant dynamics with a smaller shot budget.

\subsection{Single-band Anderson Impurity Model}\label{sec:aim}

Quantum embedding methods such as dynamical mean-field theory (DMFT)~\cite{Metzner1989,Georges1996}, density-matrix embedding theory (DMET)~\cite{Knizia2012, Knizia2013} and ghost Gutzwiller ansatz (gGut)~\cite{Lanata2017,MejutoZaera2023} map complex lattice problems onto an effective Anderson Impurity Model (AIM). In what follows, we will specifically concentrate on the single-band AIMs, obtained from converged gGut calculations. We will further use the quantum-classical algorithmic methodology presented in Ref.~\cite{GGA}, which uses the quantum selected configuration-interation (QSCI)~\cite{qsci} algorithm as the quantum impurity solver within the self-consistent gGut loop. 

In the gGut approach, a correlated impurity orbital couples self-consistently to a set of non-interacting auxiliary ``ghost'' orbitals. gGut provides a variational embedding, free of analytic-continuation and bath-discretization bottlenecks, yielding high accuracy results with only a modest number of ghost orbitals, $N_g$. The computational bottleneck of the gGut loop is the \emph{impurity solver}: at each iteration the ground state $|\Psi_{\text{GS}}\rangle$ and the one-particle RDM of the effective AIM need to be determined to update the parameters of the gGut loop. While classical exact diagonalization scales exponentially with the number of ghost orbitals, Ref.~\cite{GGA} has demonstrated that using pretrained variational local unitary cluster Jastrow (LUCJ) ansatz circuits within QSCI provides an effective quantum-assisted solver for gGut. Because gGut yields compact AIM instances with modest qubit requirements, it serves as an ideal testbed for hybrid quantum-classical solvers. The effective AIM to be solved within the self-consistent loop takes the form:
%------------------------------------------------------
\begin{align}
    &\hat{\mathcal{H}} \; = U  \hat{n}_{0\uparrow} \hat{n}_{0\downarrow} + \mu \sum_{\sigma} \hat{n}_{0\sigma} \label{eq:aim_f}\\& 
    +\sum_{a\sigma}  \Delta_{a \sigma} \left(\hat{d}^{\dagger}_{0\sigma} \hat{d}^{\phantom{\dagger}}_{a\sigma}
    + \hat{d}^{\dagger}_{a\sigma} \nonumber \hat{d}^{\phantom{\dagger}}_{0\sigma} \right)
    - \sum_{ab\sigma} \Lambda_{ba}  \hat{d}^{\dagger}_{a\sigma}\hat{d}^{\phantom{\dagger}}_{b\sigma}, \nonumber 
\end{align}
%-----------------------------------------------------
where $\hat{d}^{\dagger}_{0\sigma}$ and $\hat{d}_{0\sigma}$ are the creation and annihilation operators of the impurity orbital, while $\hat{d}^{\dagger}_{a\sigma}$ and $\hat{d}_{a\sigma}$ are the corresponding ladder operators of the ghost orbitals $a,b, \ldots \in \{1,2,\dots,N_g\}$, and the corresponding density operators are defined as $\hat{n}_{i\sigma}^{d} = \hat{d}^{\dagger}_{i\sigma} \hat{d}_{i\sigma}^{\phantom{\dagger}}$. Further, $\Delta$ is the hybridization potential, coupling the impurity orbital to ghost orbitals, and $\Lambda$ is the ghost potential, both of which are fixed by the outer gGut self-consistency loop. Crucially, the ghost orbitals can always be rotated into a basis in which they couple exclusively to the impurity orbital and not to one another, yielding a star connectivity graph:
%---------------------------------------
\begin{align}
\hat {\mathcal{H}}_{\text{star}}
\; =& U  \hat{n}_{0\uparrow} \hat{n}_{0\downarrow} + \mu \sum_{\sigma} \hat{n}_{0\sigma} 
\label{eq:AIM_star} \\
& + \sum_{a\sigma} \tilde{\Delta}_{a \sigma}\left(\hat{d}^{\dagger}_{0\sigma} \hat{\tilde{d}}^{\phantom{\dagger}}_{a\sigma}
    + \hat{\tilde{d}}^{\dagger}_{a\sigma} \nonumber \hat{d}^{\phantom{\dagger}}_{0\sigma} \right)
    - \sum_{a\sigma} \tilde{\Lambda}_{a}  \hat{\tilde{d}}^{\dagger}_{a\sigma}\hat{\tilde{d}}^{\phantom{\dagger}}_{a\sigma},\nonumber
\end{align}
%---------------------------------------
where $\hat{\tilde{d}}^{\dagger}_{a\sigma}$ ($\hat{\tilde{d}}_{a\sigma}$) are the creation (annihilation) operators of the rotated star-ghost orbitals, $\tilde{\Delta}_{a\sigma}$ denotes the effective hybridization strength, and $\tilde{\Lambda}_{a}$ represents the diagonal star-ghost energy levels.

Under the Jordan-Wigner (JW) transformation, the corresponding qubit Hamiltonian reads,
\begin{align}
\label{eq:AIM_star_JW}
\hat{\mathcal{H}}_{\mathrm{star}}^{\mathrm{JW}}
=&
\frac{\mu}{2}
\sum_{\sigma}
\left(
I-Z_{q_{0\sigma}}
\right)
\\
&+
\frac{U}{4}
\left(
I-Z_{q_{0\uparrow}}
-Z_{q_{0\downarrow}}
+Z_{q_{0\uparrow}}Z_{q_{0\downarrow}}
\right)
\nonumber\\
&-
\frac{1}{2}
\sum_{a\sigma}
\tilde{\Lambda}_{a}
\left(
I-Z_{q_{a\sigma}}
\right)
\nonumber\\
&+
\frac{1}{2}
\sum_{a\sigma}
\tilde{\Delta}_{a\sigma}
\left(
X_{q_{0\sigma}}\,
\overline{Z}^{\sigma}_{a}\,
X_{q_{a\sigma}}
+
Y_{q_{0\sigma}}\,
\overline{Z}^{\sigma}_{a}\,
Y_{q_{a\sigma}}
\right)\nonumber,
\end{align}

where $\bar{Z}^{\sigma}_{l}=\prod_{m} Z_m$ runs over the qubits strictly
between $q_{d\sigma}$ and $q_{l\sigma}$. 
We place the two impurity spin orbitals adjacent to one another at the center
of the qubit register and group the ghost orbitals according to spin. The
fermionic modes are thus ordered as
\begin{equation}
\left(
N_g\uparrow,\ldots,1\uparrow,,
0\uparrow,0\downarrow,
1\downarrow,\ldots,N_g\downarrow
\right),
\label{eq}
\end{equation}
where $(0)$ labels the impurity orbital and $(a=1,\ldots,N_g)$ label the
ghost orbitals. The total number of qubits is
$N_q=2(N_g+1)$.

Rather than relying on ground-state variational optimization, i.e. via VQE, we replace the state-preparation with time-evolved states produced through OVQRTE. The hybrid OVQRTE+QSCI workflow proceeds in two steps -- a) Subspace generation via OVQRTE wherein starting from an ansatz state $|\Psi_0\rangle$, OVQRTE executes real-time propagation under $\hat{\mathcal{H}}_{\text{star}}^{\mathrm{JW}}$. During this real-time evolution, quantum measurements sample the dominant many-body states generated along the dynamical trajectory. b) Ground-state extraction via QSCI wherein the collected configurations define a reduced, physically relevant subspace. Classical diagonalization of $\hat{\mathcal{H}}_{star}$ within this time-selected subspace yields the ground-state energy $E_{\text{GS}}$, the approximate ground-state $|\Psi_{\text{GS}}\rangle$ and the one particle reduced density matrix (1-RDM) needed for the gGut embedding update. Whilst it would, in principle, be desirable to run the full gGut self-consistency loop on quantum hardware using OVQRTE+QSCI as the impurity solver at every iteration, the accumulation of hardware and sampling noise across repeated calls could trigger uncontrolled feedback instabilities in the outer loop. Following the benchmark methodology of Ref.~\cite{GGA}, we focus on benchmarking OVQRTE+QSCI on the converged AIM instances up to 24 qubits obtained from classical gGut solutions for the Bethe lattice. This allows us to directly evaluate the accuracy of the extracted $E_{\text{GS}}$, 1-RDM, and spectral function $\mathcal{A}(\omega)$ against exact classical references.

%For the scope of this work, we therefore focus on benchmarking OVQRTE+QSCI on the converged AIM of Eq.~\eqref{eq:AIM_star} obtained from the classical gGut solution for the Bethe lattice~\cite{GGA}, directly assessing the performance as a quantum impurity solver against exact classical references. 

\begin{figure}[t]

    % Left column: ansatz + pool scaling vertically stacked

    % Right column: larger VQRTE comparison
        \centering
        \subfloat{
            \includegraphics[width=.85\linewidth]{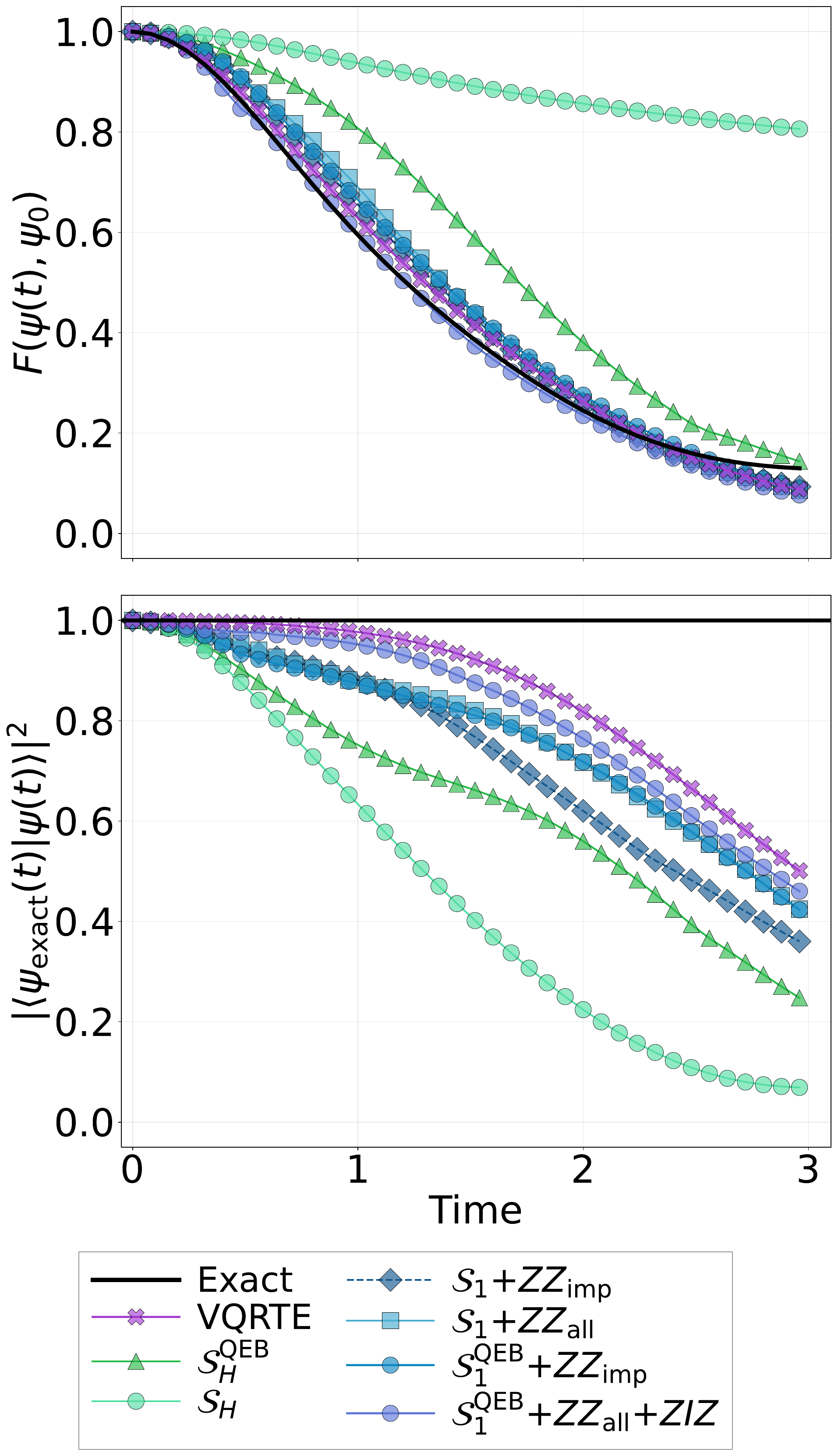}
        }

    \caption{\textbf{Simulated real-time dynamics of a $12$-qubit Anderson impurity
    model.}
    (a) Return fidelity $F(\psi(t),\psi_0)=|\langle\psi_0|\psi(t)\rangle|^2$;
    (b) fidelity to the exact state
    $|\langle\psi_{\mathrm{exact}}(t)|\psi(t)\rangle|^2$. 
    Different operator pools and their QEB variants are considered for OVQRTE, as described in the main text. VQRTE results are shown for reference. LUCJ is used as the underlying ansatz circuit for both VQRTE and OVQRTE.}
    \label{fig:12lucj}
\end{figure}

Having defined the operator pool methodology in the context of spin dynamics, we extend this selection strategy here to the fermionic operators. As a first level of approximation for a valid density matrix basis, we take the one-body reduced density
matrix (1-RDM), spanned by the hopping operators
$\hat a_p^\dagger \hat a_q$, or in Hermitian form
    $\hat a_p^\dagger \hat a_q + \hat a_q^\dagger \hat a_p$,
which describe hopping between spin-orbitals $p$ and $q$. Under
the fermion-to-qubit mapping each such operator becomes a sum of Pauli
strings, making up the operator pool we denote
$\mathcal{S}_1$, which contains $O(n^2)$ elements, where $n$ is the number of qubits. The systematic extension thereof is the
two-body reduced density matrix (2-RDM), spanned by the quartic operators
$\hat a_p^\dagger \hat a_q^\dagger \hat a_r \hat a_s$, whose count grows as
$O(n^4)$ and is costly to measure in full. Instead of the complete 2-RDM
basis, we augment $\mathcal{S}_1$ with small sets of operators that target
the leading interaction-driven correlations: (1) density-density ($ZZ$-type)
operators restricted to the impurity qubits ($\mathcal{S}_1{+}ZZ_{\rm
imp}$) or (2) extended to all nearest-neighbor pairs ($\mathcal{S}_1{+}ZZ_{\rm
all}$), which captures the diagonal (charge-charge) sector of the 2-RDM omitted by
$\mathcal{S}_1$; (3) the Pauli terms of the Hamiltonian itself
($\mathcal{S}_H$), which for this star-geometry AIM contains only $O(n)$
terms since only impurity-bath couplings are
present (4) and a next-nearest-neighbor extension of $\mathcal{S}_1{+}ZZ_{\rm
all}$ with the density-density operators of the form $Z_iI_{i+1}Z_{i+2}$. For any of these pools one may adopt a
qubit-excitation-based (QEB) variant thereof [cf.\ Eq.~\eqref{eq:QEB}], obtained by
dropping the JW parity ($Z$) strings while retaining the endpoint
operators; we mark such pools with a superscript, e.g.\ $\mathcal{S}_H^{\rm QEB}$.

When measuring operators in $\mathcal{S}_1$ (and its augmentations), the total number of required circuits
 need not equal the operator count since operators whose Pauli strings
are qubit-wise commuting (QWC) can be measured simultaneously. We summarize the scaling of selected operators in Table~
\ref{tab:pools} in \ref{app:implementation} for both, using a standard greedy QWC-grouping and 
naive measurements. The exact-JW hopping strings in $\mathcal{S}_1$ generically do not qubit-wise commute: the intermidiate $Z$-string of a longer term (e.g.,$X_1Z_2X_3$) conflicts with the $X$ or $Y$ endpoints of overlapping shorter terms (e.g. $X_1X_2$), 
%an endpoint of one term typically coincides with a qubit covered by another term's $Z$-string, and 
since $Z$ never qubit-wise commutes with $X$ or $Y$ on the same qubit. The only strings that reliably do commute are the $X\!\cdots\!X$/$Y\!\cdots\!Y$ pair belonging to the same hopping term, since they act on identical support. Grouping therefore merges only these same-term pairs, halving the circuit count but leaving the scaling $O(n^2)$. However, grouping helps dramatically for the QEB pools. For instance, stripping the $Z$-strings in the QEB
variant removes this conflict entirely, collapsing $\mathcal{S}_1^{\rm
QEB}$ to just $2$ groups (all $X$-type terms, then all $Y$-type terms)
regardless of $n$. The same pattern holds for $\mathcal{S}_H$: its raw
hybridization terms need $O(n)$ groups, while the QEB variant needs only
$3$. Diagonal ($Z$, $ZZ$) operators always commute qubit-wise and group
into a single circuit setting for any $n$; the next-nearest-neighbor
$Z_iI_{i+1}Z_{i+2}$ extension is diagonal for the same reason, so it adds no new
circuit settings beyond those already required by
$\mathcal{S}_1{+}ZZ_{\rm all}$. 
% The
% full hierarchy is
% benchmarked in Fig.~\ref{fig:12lucj} against VQRTE and exact time
% evolution $e^{-i\hat H\Delta t}$.

% \paragraph*{Local unitary cluster Jastrow ansatz.}
As the variational ansatz, we use the local unitary cluster Jastrow (LUCJ)
ansatz~\cite{motta23,GGA}, which interleaves orbital rotations with unitary
density-density (Jastrow) interactions. Details on the LUCJ ansatz can be found in Appendix~\ref{app:ansatz}. We show the cost scaling of the total number of circuit measurements per time step accounting for the number of parameters in LUCJ and the cardinality of the chosen operator set in Fig.~\ref{fig:scaling}. 
Let us now investigate the accuracy of the real-time trajectory of the LUCJ ansatz under the Hamiltonian propagator Eq.~\eqref{eq:AIM_star_JW}, approximated with OVQRTE for the operator pools listed in Table~\ref{tab:pools}.

\begin{figure}[h]
    \centering
    \includegraphics[width=0.7\columnwidth]{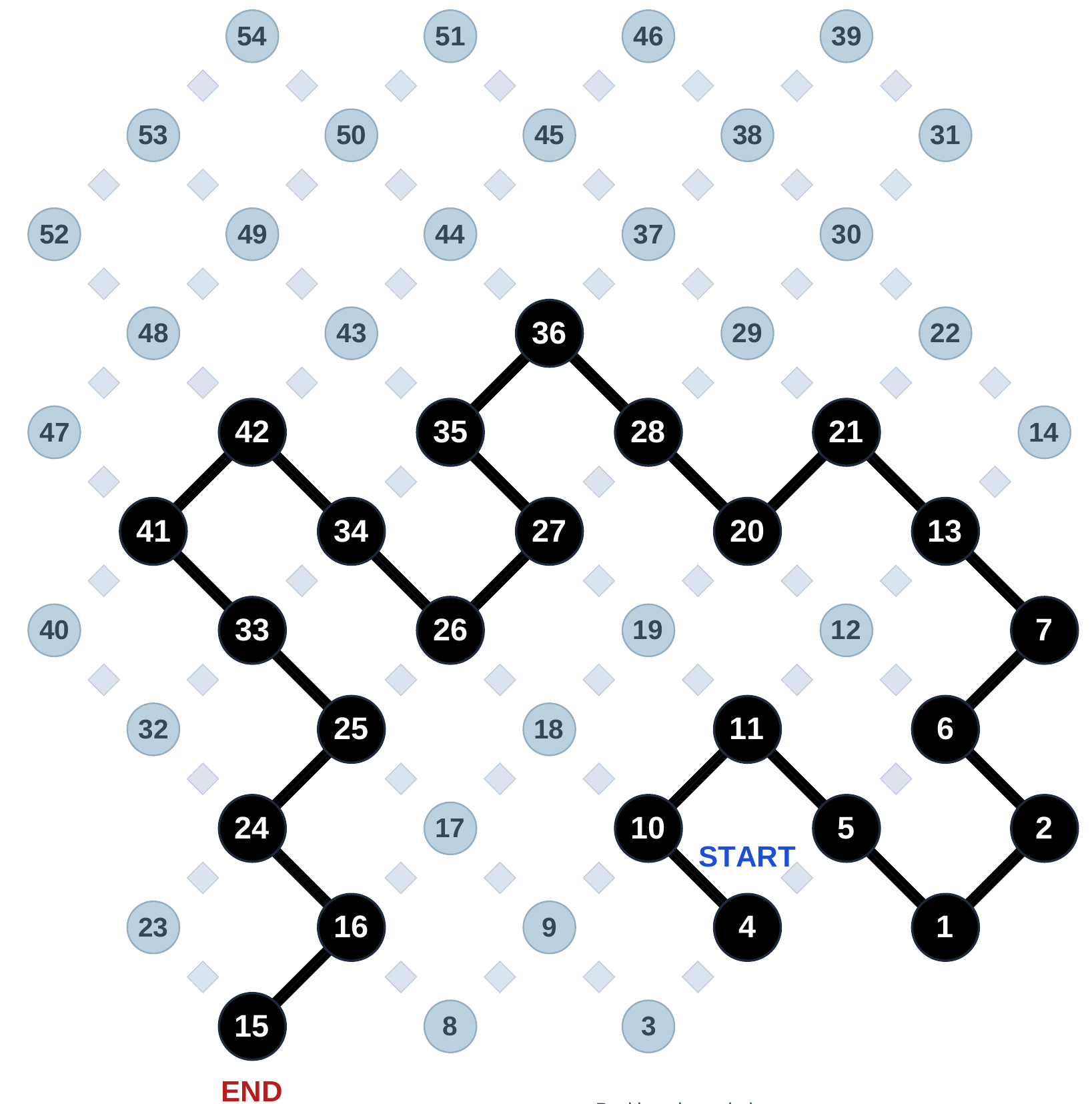}
    \caption{\textbf{Mapping of an LUCJ ansatz circuit onto the IQM Emerald
    topology.} A $24$-qubit LUCJ ansatz circuit is mapped onto a
    linear chain embedded in the device's native connectivity
    graph without the requirement for any routing via SWAP gates. The
    average calibrated CZ fidelity across the $23$ used edges is $99.71\%$.}
    \label{fig:chain-layout}
\end{figure}

\begin{figure}[h]
    \centering
     \includegraphics[width=\columnwidth]{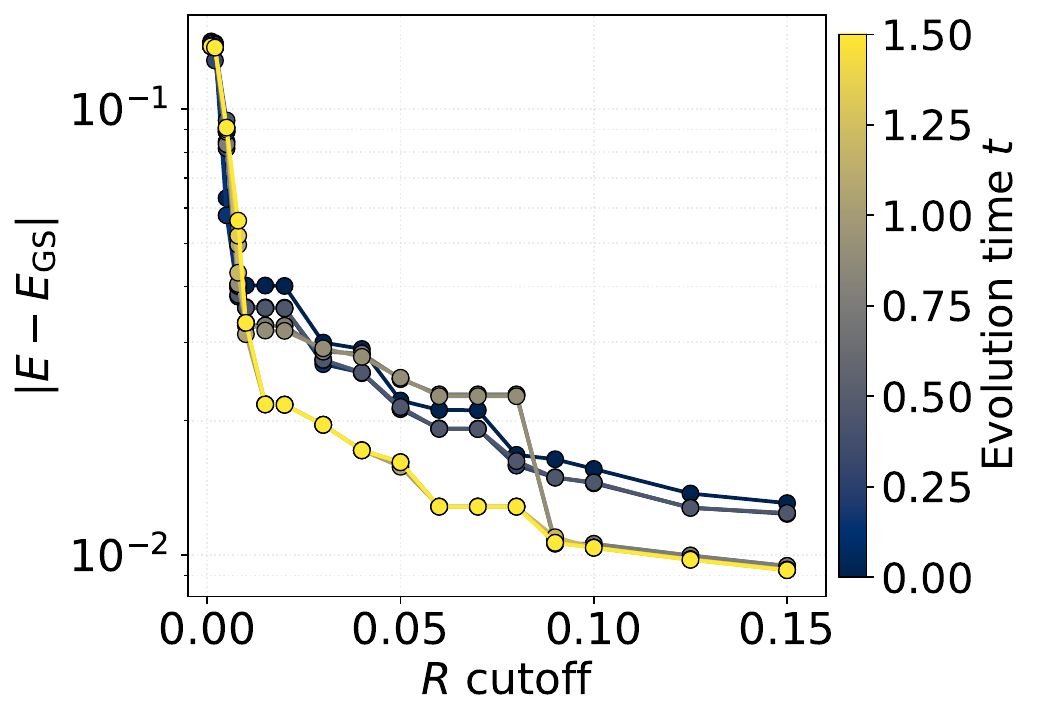}
    \caption{\textbf{OVQRTE real-time evolution under an Anderson Impurity model defined on $n=20$ qubits and with a LUCJ ansatz circuit state, executed on IQM Emerald, and followed by 
    QSCI with a configuration recovery step.} The energy error $|E-E_{\mathrm{GS}}|$ us shown as a function of the fractional subspace size $R$, color indicating the evolution
    time $t$ used. OVQRTE is run on hardware for $10$ time steps.}
    \label{fig:20qb_hw}
\end{figure}

\begin{figure*}[t]

        \centering
        
            \includegraphics[width=1.\linewidth]{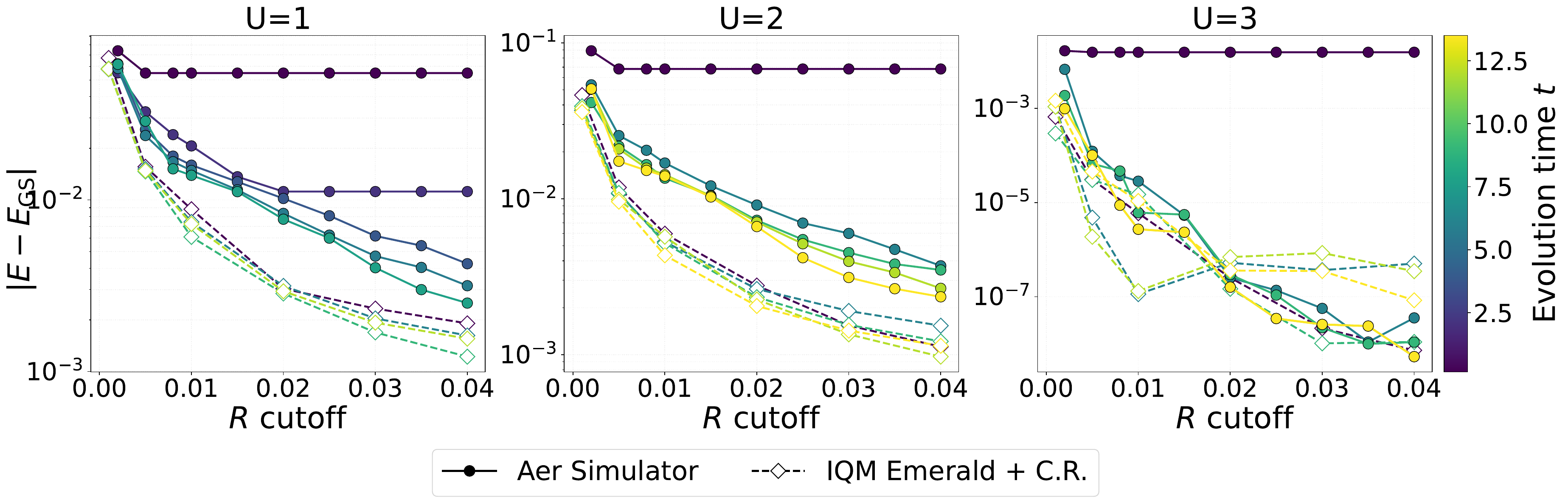}
        
\caption{\textbf{QSCI on IQM Emerald ($24$ qubits) across interaction
    strengths.}
    Energy error $|E-E_{\mathrm{GS}}|$ versus the fractional subspace size $R$, colored by
    evolution time, for $U=1,2,3$ (left to right). Solid/circles: Aer
    shot-noise simulator; dashed/squares: IQM Emerald with self-consistent
    configuration recovery~\cite{recovery}.}
    \label{fig:qsci_hw}
\end{figure*}

For the AIM defined on $n=12$ qubits, gradients are evaluated using finite differences (FD) rather than the parameter-shift rule. Because a single LUCJ variational parameter may control several Pauli-rotation gates at different circuit locations, the standard two-term parameter-shift formula would require evaluating and summing the contribution from each occurrence. FD avoids this additional bookkeeping by perturbing the shared parameter directly. The remaining OVQRTE hyperparameters, including the regularization parameter for matrix inversion and the step size, are provided in Appendix~\ref{tab:hyperparams}.

Figure~\ref{fig:12lucj} reports the return probability $F(\psi(t),\psi_0)$ (top) and the fidelity to the exact evolved state $|\langle\psi_{\rm exact}(t)|\psi(t)\rangle|^2$ (bottom), evaluated in the state-vector limit to isolate projection error from shot noise. The 1-RDM-based pools ($\mathcal{S}_1{+}ZZ_{\rm imp}$ and its QEB variant, $\mathcal{S}_1{+}ZZ_{\rm all}$) reproduce the return probability closely, but their exact-state fidelity decreases to $\sim0.3$--$0.35$ by $t=3$, showing that the good short-time approximation is not sustained over the full trajectory. The JW and QEB variants of $\mathcal{S}_1{+}ZZ_{\rm imp}$ (dashed and solid, respectively) track each other almost exactly in both panels, consistent with the comparison above: removing the JW $Z$-strings substantially reduces measurement cost without affecting the achievable accuracy.

The Hamiltonian-derived pool $\mathcal{S}_H$ performs markedly worse. Its return probability remains above $0.75$ at $t=3$, compared with the exact value of $\sim0.1$, and its exact-state fidelity decays fastest among the pools tested. Its QEB variant performs somewhat better but still trails every RDM-based pool. This is expected: $\mathcal{S}_H$ contains only the Hamiltonian's Pauli terms, rather than the 1- or 2-RDM operator content that determines the evolving observables, and is therefore insufficient to constrain the trajectory.

Adding next-nearest-neighbor density correlations markedly improves the projected evolution. In particular, $\mathcal{S}=\mathcal{S}_1^{\rm QEB}+ZZ_{\rm all}+ZIZ$ closely follows the VQRTE reference throughout the evolution and substantially reduces the deviation from the exact dynamics relative to the other OVQRTE pools. Thus, extending the projected operator space beyond one-body observables and nearest-neighbor density correlations is important for reproducing the dynamics accessible to the LUCJ ansatz.

For the fermionic mode ordering of the $n=12$ qubit AIM, the ten $Z_iZ_{i+2}$ operators are distributed symmetrically about the two impurity qubits, resolving density correlations from the immediate impurity region into the bath. This provides a possible interpretation of the improvement: the one-body sector captures the propagation of single-particle coherence, while the additional $ZZ$ and $ZIZ$ observables constrain spatially extended interaction-induced density correlations. This interpretation is empirical; the comparison establishes that the enlarged pool is sufficient to recover dynamics close to VQRTE, but does not identify which individual next-nearest-neighbor correlations are responsible.

Taken together, these results indicate that accuracy is governed less by the exactness or locality of individual operators than by whether the tracked set spans the correlations generated by the impurity-bath coupling. Matching this range, as $\mathcal{S}_1^{\rm QEB}{+}ZZ_{\rm all}{+}ZIZ$ does, closes nearly all of the remaining gap to VQRTE.

Even VQRTE's fidelity to the exact state decreases with time, reaching $\sim0.48$ by $t=3$. Since VQRTE projects the full state evolution onto the trial state manifold and therefore provides the best trajectory the manifold can support, this residual decay reflects the expressibility limit of the single-layer LUCJ ansatz rather than a projection artifact of the operator pool.

% \subsubsection{Ground-state subspace exploration via QSCI}

\begin{figure*}[t]

        \centering
        
            \includegraphics[width=0.9\linewidth]{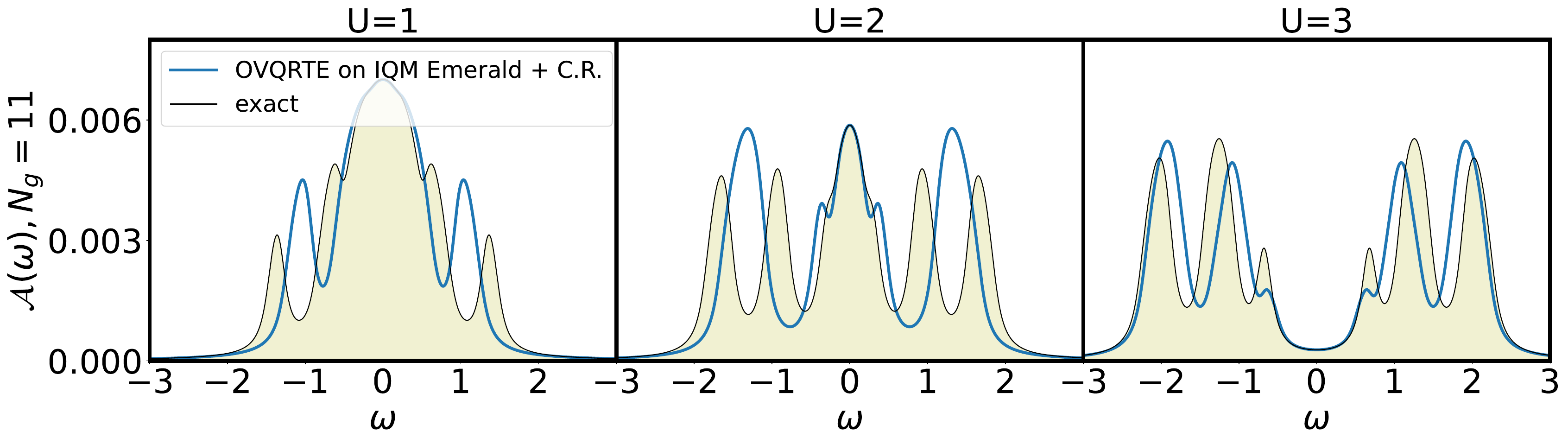}
        
\caption{\textbf{Impurity spectral function (density of states, DOS) for a $24$-qubit Anderson impurity model. }
Comparison of the exact spectral function $A(\omega)$ (yellow shaded area) with that reconstructed via OVQRTE+QSCI (blue solid line) for a $24$-qubit AIM instance at a subspace fraction of $R \approx 0.04$. Here, the circuits are selected from representative time $t$ of OVQRTE at $t=7.8,12,3.75$ for $U=1,2,3$ respectively and post-process with QSCI and S-CORE. In addition, by exploiting particle-hole and spin symmetries, OVQRTE+QSCI captures the low-frequency quasiparticle peak accurately captured, while high-frequency features reflect the effect of subspace truncation.  }
    \label{fig:dos}
\end{figure*}

%Finally, we combine OVQRTE with the quantum-selected configuration interaction (QSCI) method to explore the ground-state subspace of a $24$-qubit AIM with the LUCJ ansatz. Real-time evolution serves here as a generator of physically relevant computational-basis configurations: at each step, the evolved state is sampled in the computational basis, and the resulting bitstrings $\{|\mathbf{x}_k\rangle\}$ are collected across all sampled time steps to define a CI subspace
%\begin{equation}
%    \mathcal{S} = \text{span}\{|\mathbf{x}_k\rangle\}, \qquad
%    \hat{P}_{\mathcal{S}} = \sum_{k} |\mathbf{x}_k\rangle\langle\mathbf{x}_k|.
%    \label{eq:subspace}
%\end{equation}
%The Hamiltonian is then diagonalized within it,
%\begin{equation}
%    \hat{P}_{\mathcal{S}} \hat{\mathcal{H}}^{\text{emb}} \hat{P}_{\mathcal{S}} \, |\Psi^{\mathcal{S}}_{\text{GS}}\rangle = E_{\text{GS}} \, |\Psi^{\mathcal{S}}_{\text{GS}}\rangle,
%    \label{eq:diag}
%\end{equation}
%yielding an approximate ground state $|\Psi^{\mathcal{S}}_{\text{GS}}\rangle$ and energy $E_{\text{GS}}$ restricted to the sampled subspace $\mathcal{S}$. 

% For $n=20$ OVQRTE training,  we use a step size of $\Delta t=0.15$, gather 
Next, we combine OVQRTE with QSCI to explore the ground-state subspace of the AIM using the LUCJ ansatz for $n=\{20,24\}$ qubits. This corresponds to  $N_g=\{9,11\}$ ghost orbitals, respectively. we verify that the qubit ordering maps efficiently onto the native
connectivity of the device. Figure~\ref{fig:chain-layout} shows the
resulting $24$-qubit linear chain embedded in the IQM Emerald topology.
All $23$ consecutive pairs in the chain coincide with native two-qubit
gates on the device, with an average calibrated fidelity of $99.71\%$
across these edges. Because the LUCJ circuit's orbital-rotation
(Givens/$XX{+}YY$) and Jastrow layers act only between adjacent qubits
in this same ordering, every two-qubit gate in the circuit executes
directly as a native gate, without additional SWAP routing. The quantum circuit for $n=20,24$ qubits transpiles into a total of $398,574$ native CZ gates on the IQM quantum hardware, corresponding to a circuit depth of $134,158$ respectively.\par
We initialize LUCJ ansatz with random parameters and execute the OVQRTE-driven QSCI calculations entirely on IQM Emerald at $n=20$ qubits
(Fig.~\ref{fig:20qb_hw}). For this experiment, we used typical readout error mitigation \cite{m3_readout} where qubits were measured and calibrated to build an error profile, which was then used to correct the raw measurement results. At each of several evolution times $t\leq 1.5$, the gradients are obtained by SPSA combined with moment averaging, which we describe in more detail in Appendix~\ref{app:implementation}. 
We use $N_{\text{sh}}=10^3$ shots
for the gradient and evolution vectors of Eq.~\eqref{eq:app-ovqrte-b}, $N_{\text{sh}}=10^4$ for subspace sampling and take $N_s=26$ SPSA samples for averaging (see Appendix~\ref{app:spsa}) all per time step. Further, we select the operator pool to be $\mathcal{S}=\mathcal{S}_1+ZZ$. At each time step, the
OVQRTE-evolved LUCJ state is sampled on the quantum hardware, the resulting bitstrings define a CI subspace of size $|\chi|$, and we define the cutoff as $R=|\chi|/\text{dim}(H)$ where $\text{dim}(H)=\binom{L}{L/2}^{2}$, with $L$ being the number of impurity+bath orbitals at half filling. The Hamiltonian is diagonalized within this subspace to obtain the approximate ground state energy of the Hamiltonian. The
subspace-construction procedure (sampling protocol, subspace
assembly, and configuration recovery) is held fixed across all times
shown, so that $0\leq t \leq1.5$ is the only quantity varied between curves, as indicated through color.

The $20$-qubit result shows that the time-evolution explores increasingly relevant
configurations even under hardware noise. In Fig.~\ref{fig:20qb_hw}, the
energy error plateaus around $|E-E_{\rm GS}|\sim10^{-1}$ at the initial time ($t{=}0$), once
$R\gtrsim0.01$: enlarging the subspace further adds little,
since samples from the untrained LUCJ state carry limited information
about the ground state. As the state is evolved, the achievable energy
error improves substantially at every fixed subspace size: intermediate
times ($t\lesssim0.5$) reach $|E-E_{\rm GS}|\sim3\text{--}4\times10^{-2}$,
and later times ($t\gtrsim1.0$) continue downward to
$|E-E_{\rm GS}|\approx10^{-2}$ when setting $R=0.15$. Because the
subspace-construction procedure is identical at every $t$, this improvement can be attributed entirely
to the evolution itself: later times generate a set of sampled
bitstrings that is more relevant to the ground state than earlier ones. 
The corresponding simulator-side training
trajectory is given in Appendix~\ref{app:20qb-training} for direct
comparison against the hardware result presented here.

We next test whether this behavior
persists to larger system size and across different interaction strengths. We scale
the same OVQRTE-QSCI pipeline to $24$ qubits and execute it directly on
IQM Emerald at three interaction strengths, $U=\{1,2,3\}$
(Fig.~\ref{fig:qsci_hw}).

Along the simulator training trajectory, we
select several representative evolution times $t=\{0, 2, 3.75, 5.7, 7.8\}$ for $U=1$ and  $t\in \{0,3.75,6,9,12\}$ for $U=2,3$ spanning early, intermediate, and late stages of the evolution and run the
corresponding circuits on the quantum hardware, sampling bitstrings that are fed into
QSCI with self-consistent configuration recovery~\cite{recovery}; the same procedure is repeated on the
simulator with shot-noise at the same set of times $t$ and subspace fractions $R$ for
direct comparison.

At the earliest evolution time in each panel ($t\approx2$, darkest
curves), the energy error plateaus at a comparatively large value
independent of $R$ and essentially independent of the hardware versus
simulator choice. The same signature was seen at $20$ qubits: an insufficiently
evolved state does not populate ground-state-relevant configurations,
regardless of how it is sampled. As $t$ increases, both hardware and
simulator curves descend substantially, confirming that the mechanism
generating useful configurations is the evolution itself, not just the
sampling contribution from hardware and recovery.

At weak and moderate coupling ($U=
{1,2
}$), the quantum hardware with configuration
recovery matches or outperforms the shot-noise simulator across the full
range of $R$ tested, consistent with the noise-assisted-exploration
picture described above: hardware sampling draws from a broader
configuration distribution than idealized shot noise, and recovery
fixes the unphysical additions while retaining the useful ones, net
enlarging the effective subspace. 

At strong coupling ($U=3$), this
advantage holds only up to moderate $R$: aside from the
unevolved state at $t=0$, which plateaus near $|E-E_{\rm GS}|\sim2\times10^{-2}$
regardless of subspace size, every evolution time reaches machine precision
together once $R\gtrsim0.02$, with hardware and simulator
curves overlapping closely throughout. Little separates early and late
evolution times here whereby at strong coupling the ground-state wavefunction
concentrates most of its weight on comparatively few configurations, so a modestly evolved state already samples
enough of the relevant bitstrings, and further evolution has little
additional room to yield gains.

Beyond ground-state energies, the wavefunctions extracted from OVQRTE+QSCI enable the evaluation of dynamic correlation functions, such as the impurity spectral function (density of states, DOS) $\mathcal{A}(\omega)= -\frac{1}{\pi}\mathrm{Im}\,G(\omega)$ where $G(\omega)$ is the impurity Green's function evaluated at frequency $\omega$, across interaction strengths $U \in \{1,2,3\}$ (Fig.~\ref{fig:dos}). Following the gGut embedding formalism~\cite{GGA}, the local Green's function $G(\omega)$ is obtained from a closed-form analytical expression parameterized directly by the one-particle reduced density matrix (1-RDM) $\rho = \bra{\Psi_{\text{GS}}} \hat{d}^{\dagger} \hat{d}^{\phantom{\dagger}}  \ket{\Psi_{\text{GS}}}$ evaluated on the ground state $|\Psi_{\text{GS}}\rangle$ [cf.\ Eq.~(12) of Ref.~\cite{GGA}]. Here, the 1-RDM is extracted from the approximate QSCI ground state within the symmetry subspace $\mathcal{C}$ at a $R \approx 0.04$, using representative OVQRTE-evolved circuits at evolution times $t=7.8, 12.5,$ and $3.75$ for $U=1,2,3$, respectively. At weak and moderate couplings ($U=1,2$), OVQRTE+QSCI (blue solid line) closely matches the exact reference (yellow shaded area) at a compact subspace fraction of $R \approx 0.04$ (enforcing particle-hole and spin symmetries), faithfully reproducing the central quasiparticle peaks. The analysis is particularly notable at $U=3$, which marks the regime near the transition to a Mott-insulating phase. Parallel to the high energy resolution achieved at $U=3$ in Fig.~\ref{fig:qsci_hw}, the reconstructed DOS is almost fully resolved here as well: OVQRTE+QSCI correctly captures the opening of the Mott gap around the Fermi level ($\omega = 0$). Minor discrepancies remain only in the precise energetic location and line shape of the high-frequency Hubbard side peaks, which require expanding the CI subspace to pin down completely, while the essential Mott physics and spectral gap are already well established.

\section{Conclusion}
\label{sec:conclusions}

We have introduced a variational method for real-time quantum dynamics that
replaces the full McLachlan principle with its projection onto a chosen manifold
of operators. Rather than enforcing the equation of motion across the entire
state space, OVQRTE tracks only a polynomially large set of physically relevant
observables and fits the parameter velocity to reproduce their exact Ehrenfest
dynamics. This projection turns the size of the operator set into a tunable knob
that trades accuracy against circuit depth and measurement cost, and it recovers
standard VQRTE in the limit of a complete set.

For local spin systems the trade-off is strongly favorable. On the Heisenberg
model, tracking only two-local nearest-neighbor observables reproduces the exact
trajectory, scales far more gently with system size than VQRTE, and is markedly
more resilient to shot noise, since it requires only expectation values rather
than quantum metric tensor and gradient evaluations. For fermionic problems,
even where the dynamics are only approximate, the states visited along the
trajectory are useful in their own right. Applying OVQRTE as a generator of
computational-basis configurations, we showed that the subspace accumulated over
successive time steps recovers a good fraction of the ground-state manifold of a
$24$-qubit Anderson impurity model through QSCI. Executed on IQM Emerald, the
subspace method benefited from hardware sampling: the device produces a broader
set of unique bitstrings than the idealized sampler, and self-consistent
configuration recovery turned this diversity into a systematic improvement in
the recovered energy.

Taken together, these results establish operator-projected dynamics as a
low-cost, hardware-friendly primitive. As devices approach the early
fault-tolerant regime, it can serve as a subroutine for quantum-chemistry tasks
beyond the dynamics itself such as generating subspaces for ground-state energies, as
demonstrated here, or for dynamical quantities such as Green's functions and
spectral functions, where real-time correlators are the natural output. For
fermionic problems, faithfully capturing correlated dynamics can require larger
operator sets, which may become a bottleneck at scale. A natural direction for
future work is therefore to prune the operator set adaptively, for instance by
weighting the residual and discarding the operators that contribute least to the
parameter update. We view the present work as a proof of concept that maps out
the method and its trade-offs. The flexibility in choosing the operator set, the
ansatz, and the downstream application leaves considerable room to explore, and
we hope it invites further study on larger and more challenging systems.

\section{Acknowlegdements}
The authors thank Carlos Benavides-Riveros, Prachi Sharma, Alessio Calzona, Manuel Algaba and Christian Mendl, for useful discussions. We acknowledge the use of Qiskit and Qiskit Algorithms for the numerical simulations\cite{qiskit-varqrte}.
\bibliographystyle{apsrev4-2}
\bibliography{bib}

\clearpage
\onecolumngrid
\section{APPENDIX}

\subsection{Method of implementation}\label{app:implementation}

\subsubsection{Parameter-update linear systems}\label{app:linsys}

Both OVQRTE and VQRTE determine the parameter update $\dot{\bm\theta}$
from a linear system of the same abstract form,
\begin{equation}
    K\,\dot{\bm\theta} = \bm b,
    \label{eq:app-generic-linsys}
\end{equation}
obtained by least-squares matching of a target flow in parameter space; $K$
and $\bm b$ differ between the two methods. For OVQRTE,
\begin{align}
    K &= A = G^\top G, &
    G_{\alpha j} &= \partial_{\theta_j}\langle\hat O_\alpha\rangle,
    \label{eq:app-ovqrte-K}\\
    \bm b &= C, &
    C_j &= \sum_{\alpha=1}^{|\mathcal{S}|} G_{\alpha j}\, i\langle[\hat H,\hat O_\alpha]\rangle;
    \label{eq:app-ovqrte-b}
\end{align}
for VQRTE,
\begin{align}
    K &= Q, &
    Q_{ij} &= \mathrm{Re}\bigl[\langle\partial_i\psi|\partial_j\psi\rangle
              -\langle\partial_i\psi|\psi\rangle\langle\psi|\partial_j\psi\rangle\bigr],
    \label{eq:app-VQRTE-K}\\
    \bm b &= V, &
    V_k &= \mathrm{Im}\bigl[\langle\partial_k\psi|(\hat H-E)|\psi\rangle\bigr],
    \quad E=\langle\psi|\hat H|\psi\rangle.
    \label{eq:app-VQRTE-b}
\end{align}
$G\in\mathbb{R}^{M\times p}$ is the observable Jacobian of Sec.~\ref{sec:theory};
$Q\in\mathbb{R}^{p\times p}$ is the quantum geometric tensor (QGT).
 Every entry of $G$ is
the derivative of an expectation value of a fixed Hermitian operator
$\hat O_\alpha$ on a single state $|\psi(\bm\theta)\rangle$, and is
therefore obtainable from ordinary expectation-value measurements, with no
ancilla. Every entry of $Q$ is an inner product between two different
tangent states, $|\partial_i\psi\rangle$ and $|\partial_j\psi\rangle$ which is an
overlap, rather than an expectation value on a single state. 
For VQRTE, we use the real-time McLachlan implementation provided by
Qiskit~\cite{qiskit-varqrte}, which evaluates the QGT $Q$ and evolution vector
$\bm V$ in Eqs.~\eqref{eq}--\eqref{eq} using
linear-combination-of-unitaries (LCU) gradient circuits. The LCU construction
uses an ancilla qubit to coherently interfere the circuit contributions
associated with two parameter derivatives, such that the required real or
imaginary part of the corresponding overlap is obtained from an ancilla
expectation value. In this way, $Q_{ij}$ is evaluated from pairwise tangent-state
overlaps, while $V_k$ is obtained from overlaps involving a single tangent
direction and the Hamiltonian.
For an ansatz with $p$ variational parameters, evaluation of the QGT requires
all parameter pairs $(i,j)$ and therefore has $O(p^2)$ circuit-setting
complexity (only the independent triangular part need be evaluated), whereas
the evolution vector contains $p$ components and scales as $O(p)$. The QGT
therefore determines the leading measurement-cost scaling of VQRTE. Accordingly,
when comparing measurement costs below, we use $C_{\mathrm{VQRTE}} \sim p^2$
as the VQRTE circuit-setting scaling, omitting constant prefactors and
subleading $O(p)$ contributions.\\
For ansätze composed of Pauli-rotation gates, $G_{\alpha j}$ is obtained
exactly via the two-term parameter-shift rule (PSR)~\cite{grad1},
\begin{equation}
    G_{\alpha j}(\bm\theta) = \frac{1}{2\sin s}
    \Bigl(\langle\hat O_\alpha\rangle_{\bm\theta+s\vec e_j}
         -\langle\hat O_\alpha\rangle_{\bm\theta-s\vec e_j}\Bigr),
    \label{eqn:psr}
\end{equation}
requiring $O(p)$ circuit settings for each $\hat O_\alpha\in\mathcal{S}$
(measured simultaneously across $\alpha$ via the QWC grouping). However, when the ansatz does not satisfy the PSR eigenvalue condition (e.g.\ LUCJ,
Sec.~\ref{sec:aim}), $G_{\alpha j}$ is instead estimated by central finite
differences with step $h$,
\begin{equation}
    G_{\alpha j}(\bm\theta) \approx
    \frac{\langle\hat O_\alpha\rangle_{\bm\theta+h\vec e_j}
         -\langle\hat O_\alpha\rangle_{\bm\theta-h\vec e_j}}{2h},
    \label{eqn:fd}
\end{equation}
at the same $O(p)$ circuit-setting cost.

\paragraph*{Stochastic (SPSA) gradient estimation}\label{app:spsa}

Both linear systems can alternatively be assembled with $O(1)$
circuit-setting cost per sample using the simultaneous perturbation
stochastic approximation (SPSA), at the expense of variance. Sampling $k$ perturbation directions
$\bm\Delta_l\in\{-1,1\}^p$ uniformly at random,
\begin{equation}
    \widetilde G_\alpha(\bm\theta) = \frac{1}{k}\sum_{l=1}^k
    \frac{\langle\hat O_\alpha\rangle_{\bm\theta+c\bm\Delta_l}
         -\langle\hat O_\alpha\rangle_{\bm\theta-c\bm\Delta_l}}{c\,\bm\Delta_l}
    \in\mathbb{R}^p,
    \label{eqn:spsa-ovqrte}
\end{equation}
for perturbation magnitude $c>0$, held fixed rather than annealed. The
linear system~\eqref{eq:app-generic-linsys} is then assembled with
$\widetilde A = \widetilde G^\top\widetilde G$; since
$\mathbb{E}[(\bm a^\top\bm\Delta)\bm\Delta]=\bm a$ for i.i.d.\ Rademacher
$\bm\Delta$, $\mathbb{E}[\widetilde A]=A$. The cost is
$O(k\,C_{\mathrm{pool}})$, independent of $p$.

For VQRTE, the analogous unbiased estimators for $Q$ and $V$,
using nested and single perturbation directions respectively, follow
Refs.~\cite{gacon2021,gacon2023}, at $O(1)$ circuit settings per sample,
independent of $p$. In this work, we do not implement or benchmark this variant and include the reference
only for completeness alongside the OVQRTE estimator above.

\paragraph*{Moment averaging.}
A single SPSA sample (or batch average) is a poor estimator of the
instantaneous gradient during time evolution, since a plain running average
over all past steps cannot track a time-dependent target. Following the
momentum-based averaging scheme used for variational time evolution in
Ref.~\cite{gacon2023}, we instead combine estimates across steps with an
exponential moving average,
\begin{equation}
    \overline{G}_i^{\,t} = \tau\,\overline{G}_i^{\,t-1} + (1-\tau)\,\widetilde G_i^{\,t},
    \label{eqn:moment-avg}
\end{equation}
and analogously for $C$ (OVQRTE) or $Q,V$ (VQRTE), with momentum
$\tau\in(0,1)$. We do not derive an optimality guarantee for this choice;
we report it as the estimator used in our runs, consistent with its
prior empirical use for stochastic imaginary-time evolution~\cite{gacon2023}.

Table.~\ref{tab:hyperparams} lists the regularization parameter ($\lambda$), time step / learning
rate ($\Delta t$), finite-difference step ($h_\mathrm{FD}$), SPSA
perturbation amplitude ($c_k$), momentum decay ($\tau$), and number of samples
($N_s$) for each configuration shown in the figures.
Dashes (\textbf{--}) indicate the parameter is not applicable for that
configuration (e.g.\ no SPSA is used, or the gradient method does not employ a
finite-difference step).

\begin{table*}[h!]
\centering
\begin{tabular}{@{}lllllll@{}}
\toprule
Configuration
  & $\lambda$
  & $\Delta t$
  & $h_\mathrm{FD}$
  & $c_k$
  & $\tau$
  & $N_s$ \\
\midrule
\multicolumn{7}{@{}l}{\textit{Fig.~\ref{fig:heisenberg} \quad Heisenberg 12qb}} \\[2pt]
\quad VQRTE (SV)                         & $10^{-5}$ & 0.07 & --- & --- & --- & --- \\
\quad VQRTE ($N_\mathrm{sh}=10^{4}$)     & $10^{-1}$ & 0.07 & --- & --- & --- & --- \\
\quad VQRTE ($N_\mathrm{sh}=10^{5}$)     & $10^{-2}$ & 0.07 & --- & --- & --- & --- \\
\quad OVQRTE $\mathcal{S}_{2\mathrm{NN}}$ (SV)      & $10^{-5}$ & 0.07 & --- & --- & --- & --- \\
\quad OVQRTE ($N_\mathrm{sh}=10^{4}$)    & $10^{-2}$ & 0.07 & --- & --- & --- & --- \\
\quad OVQRTE ($N_\mathrm{sh}=10^{5}$)    & $10^{-2}$ & 0.07 & --- & --- & --- & --- \\
\midrule
\multicolumn{7}{@{}l}{\textit{Fig.~\ref{fig:12lucj} \quad AIM 12qb --- pool comparison}} \\[2pt]
\quad VQRTE ($n=12$)   & $10^{-4}$ & 0.08 & ---   & --- & --- & --- \\
\quad OVQRTE  ($n=12$)   & $10^{-5}$ & 0.08 & $10^{-2}$ & --- & --- & --- \\
\midrule
\multicolumn{7}{@{}l}{\textit{Fig.~\ref{fig:20qb_hw} \quad AIM 20qb --- hardware + simulator}} \\[2pt]
\quad Hardware (IQM Emerald + S-CORE) & $10^{-3}$ & 0.15 & $10^{-2}$ & 0.1 & 0.9 & 22 \\
\quad Simulator (Aer)                  & $10^{-3}$ & 0.15 & $10^{-2}$ & 0.1 & 0.9 & 22 \\
\midrule
\multicolumn{7}{@{}l}{\textit{Fig.~\ref{fig:qsci_hw} \quad AIM 24qb, $U/t = 1,2,3$}} \\[2pt]
\quad $n=24$, $U/t=1$ & $10^{-2}$ & 0.15 & $10^{-2}$ & 0.1 & 0.9 & 24 \\
\quad $n=24$, $U/t=2$ & $10^{-2}$ & 0.15 & $10^{-2}$ & 0.1 & 0.9 & 24 \\
\quad $n=24$, $U/t=3$ & $10^{-2}$ & 0.15 & $10^{-2}$ & 0.1 & 0.9 & 24 \\
\bottomrule
\end{tabular}
\caption{Hyperparameters for all paper figures.
$\lambda$: regularisation (lstsq rcond or ridge penalty).
$\Delta t$: time step / learning rate.
$h_\mathrm{FD}$: finite-difference perturbation step.
$c_k$, $\tau$, $N_s$: SPSA perturbation amplitude, momentum decay, and number of
gradient samples.
Dashes indicate the parameter is not applicable.}
\label{tab:hyperparams}
\end{table*}
\FloatBarrier

\subsubsection{Variational Ansatz}\label{app:ansatz}
\begin{figure*}[t]

        \centering
        \subfloat[\label{fig:hea}]{
            \includegraphics[width=.45\linewidth]{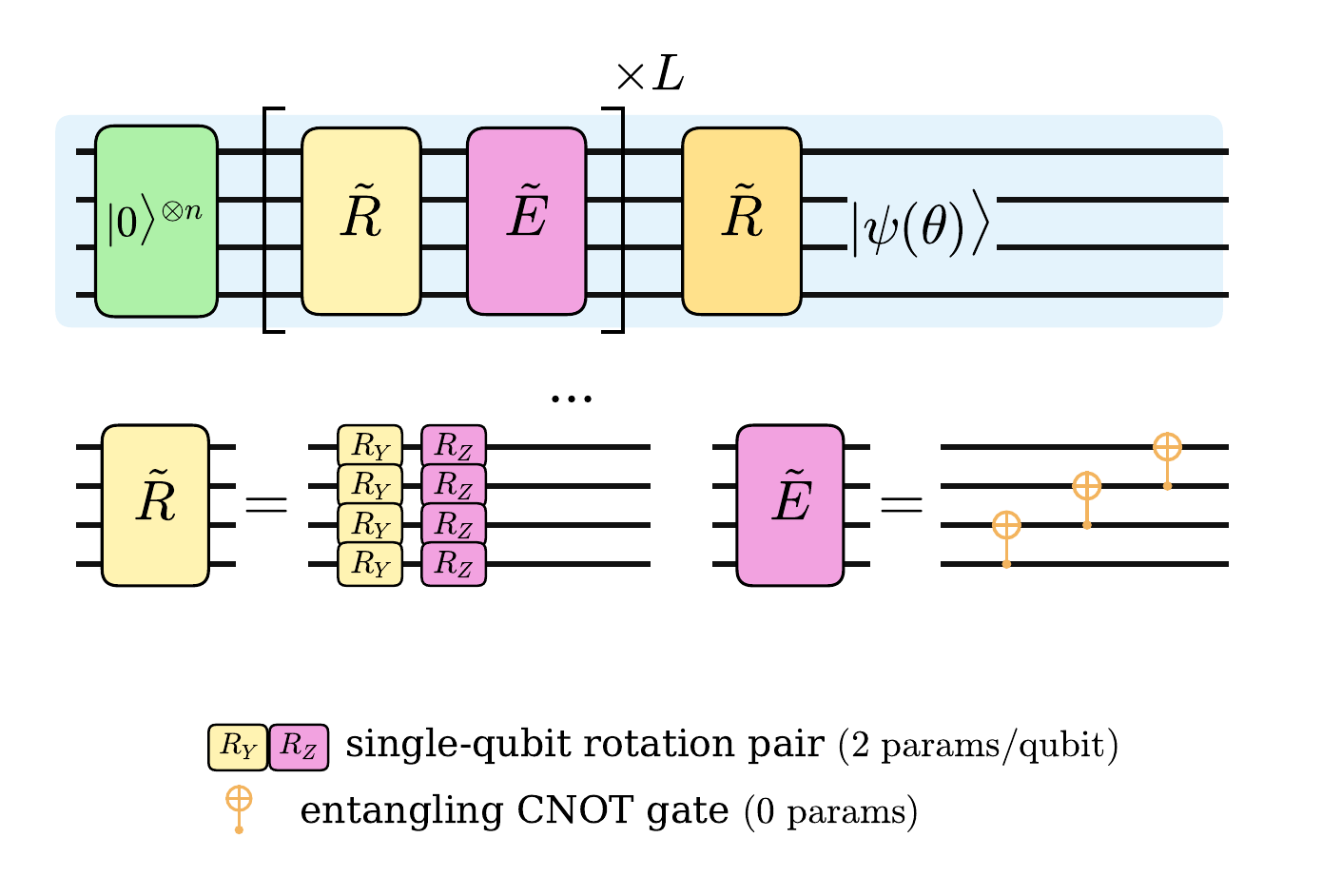}
        }
        \hfill
        \subfloat[\label{fig:gf}]{
            \includegraphics[width=0.5\linewidth]{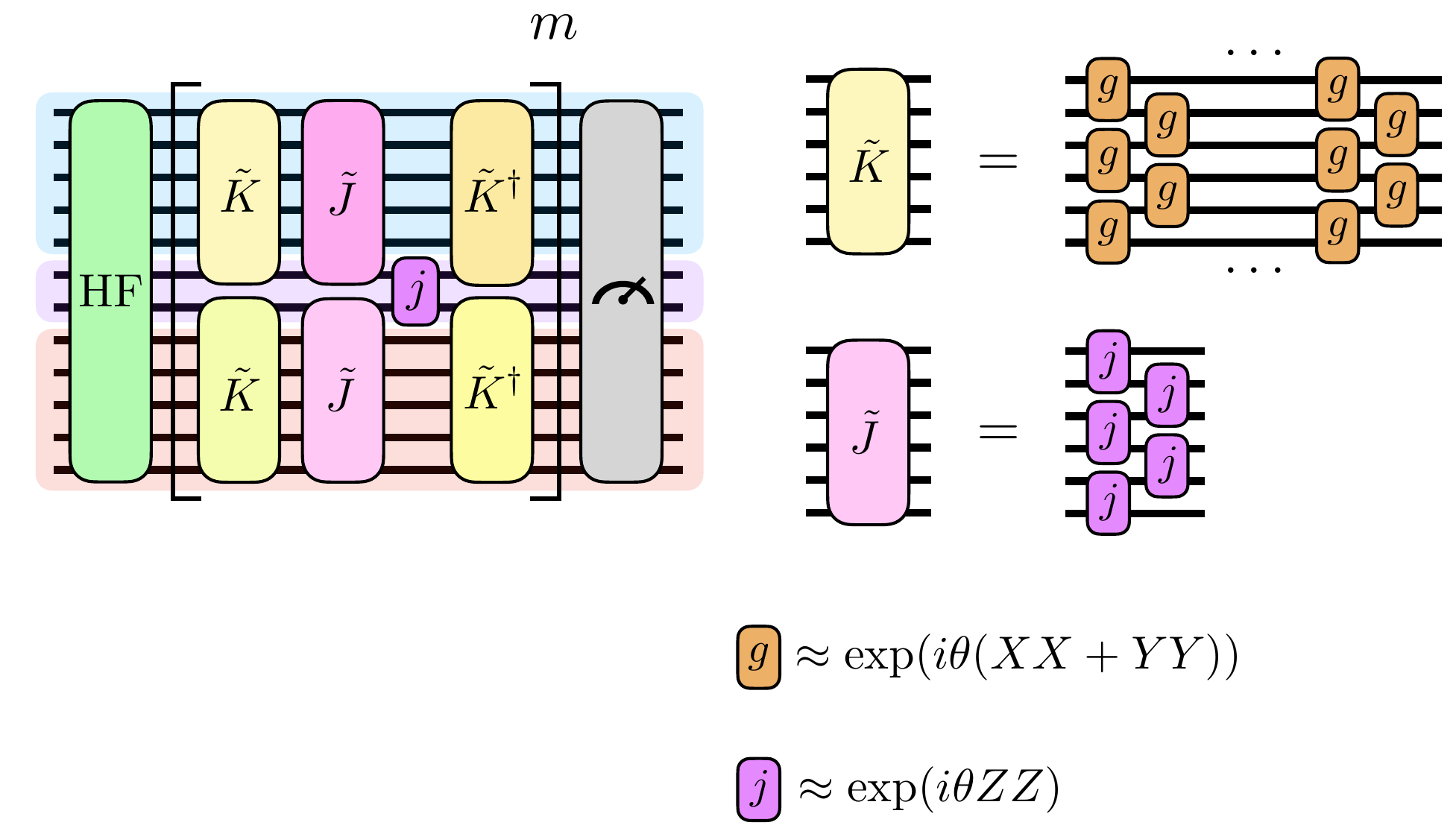}
        }
\caption{%
Variational ansatz circuits compared in this work.
\textbf{(a)} Hardware-efficient ansatz (HEA) for the Heisenberg model:
the dashed-box block ($R_Y(\theta)R_Z(\theta)$ on every qubit, $2n$
parameters, followed by a fixed CNOT entangling ladder, 0 parameters)
repeats $L$ times as indicated by the brace; the circuit is closed
with one additional single-qubit rotation layer so the last
entangling layer is not immediately followed by measurement. For
$n = 4$ qubits this gives $2n(L+1) = 8(L+1)$ parameters in total.
\textbf{(b)} LUCJ ansatz for the Anderson impurity model: both
$\tilde K$ and $\tilde J$ are realized as the same linear brick-wall
network of nearest-neighbor two-qubit gates ($g \approx
\exp(i\theta(XX+YY))$ for $\tilde K$, $j \approx \exp(i\phi\,ZZ)$ for
$\tilde J$): alternating columns of $(0,1),(2,3)$ then $(1,2)$ gates
give full pairwise connectivity among $n_o$ orbitals in $n_o$ layers,
using $n_o(n_o-1)/2$ gates (and independent parameters) total, shared
between the $\uparrow$ and $\downarrow$ sectors. }
    \label{fig:ansatz}
\end{figure*}

Starting from a Hartree-Fock reference state
$|\Psi_0\rangle$, one LUCJ layer applies
\begin{equation}
    |\Psi_{\mathrm{LUCJ}}\rangle
    = \prod_{m=1}^{L} e^{i\hat K_m}\, e^{i\hat J_m}\, e^{-i\hat K_m}\,|\Psi_0\rangle ,
    \label{eq:LUCJ}
\end{equation}
where $\hat K_m=\sum_{ i<j,\sigma}\kappa^{m\sigma}_{ij}\,
\hat d_{i\sigma}^\dagger \hat d_{j\sigma}+\mathrm{h.c.}$ is the orbital rotation and
\[
\hat J_m
=
\gamma^{m\uparrow\downarrow}
\hat n_{0\uparrow}\hat n_{0\downarrow}
+
\sum_{\langle i,j\rangle,\sigma}
\gamma^{m\sigma}_{ij}\,
\hat n_{i\sigma}\hat n_{j\sigma},
\qquad
\hat n_{i\sigma}
=
\hat d_{i\sigma}^\dagger \hat d_{i\sigma},
\]
is the local density-density generator. Here, $\langle i,j\rangle$
denotes nearest-neighbor orbitals within each spin sector. Same-spin
density-density correlations are included between neighboring orbitals,
whereas the opposite-spin interaction is restricted to the impurity
orbital, consistent with the interaction structure of $\hat H$.

% \paragraph*{Circuit realization.}
Each orbital rotation $e^{i\hat K_m}$ is a Gaussian, particle-number-conserving
unitary and can be decomposed into a network of nearest-neighbor orbital
rotations. Under the Jordan-Wigner (JW) transformation, the two-mode rotations
are implemented by gates of the form
\begin{equation}
    \exp\!\left[
        i\theta_{ij}^{m\sigma}
        \left(
            X_{i\sigma}X_{j\sigma}
            +
            Y_{i\sigma}Y_{j\sigma}
        \right)
    \right],
\end{equation}
together with single-qubit $R_Z$ rotations of the form
\begin{equation}
    \exp\!\left(i\lambda_{i}^{m\sigma} Z_{i\sigma}\right),
\end{equation}
which account for the single-particle phase degrees of freedom of the
orbital rotation. The parameters $\theta_{ij}^{m\sigma}$ and
$\lambda_i^{m\sigma}$ are determined by the coefficients
$\kappa_{ij}^{m\sigma}$ of $\hat K_m$.

The Jastrow layer $e^{i\hat J_m}$ is represented
by nearest-neighbor $ZZ$ phase rotations within each spin sector,
\begin{equation}
    \prod_{\langle i,j\rangle,\sigma}
    \exp\!\left(
        i\phi_{ij}^{m\sigma}
        Z_{i\sigma}Z_{j\sigma}
    \right),
\end{equation}
together with a single cross-spin interaction between the two impurity
orbitals,
\begin{equation}
    \exp\!\left(
        i\phi_{0}^{m\uparrow\downarrow}
        Z_{0\uparrow}Z_{0\downarrow}
    \right).
\end{equation}
Since the same-spin $ZZ$ terms act only on nearest-neighbor pairs and
mutually commute, they can be arranged in two parallel sublayers, while
the impurity $Z_{0\uparrow}Z_{0\downarrow}$ rotation can be executed in
parallel with the remaining Jastrow terms. The corresponding circuit is shown in can be found in  \ref{fig:ansatz}.

The essential feature of Eq.~\eqref{eq:LUCJ} is that the Jastrow operator is
sandwiched between two mutually inverse orbital rotations,
$e^{i\hat K_m}$ and $e^{-i\hat K_m}$. The two rotations therefore do not
constitute independent variational blocks: a single set of parameters
defining $\hat K_m$ determines both, with the conjugate rotation obtained by
reversing their signs. Consequently, only one set of orbital-rotation
parameters, together with the parameters of the diagonal $\hat J_m$, is
optimized per layer. This reduces the number of orbital rotation parameters
by a factor of two compared with an ansatz containing two independent
orbital-rotation blocks. 

Following Ref.~\cite{GGA}, the number of optimized parameters then scales as
$O\!\bigl(L\,N_q^2\bigr)$, where $L$ denotes the number of repeated LUCJ layers in Eq. \ref{eq:LUCJ} and $N_q$ is the total number of spin-orbitals, equal to the number of qubits, $n$ after JW mapping.

\subsubsection{Hierarchy of pool scaling}
% ============================================================
%  Text following the "Hierarchy of pool scaling" table.
%  Two-column safe.
% ============================================================

Table~\ref{tab:pools} collects the operator pools used in the $12$-qubit
LUCJ AIM experiment of Fig.~\ref{fig:12lucj}, where each is benchmarked
against VQRTE and exact time evolution, alongside
$\mathcal{S}_{2\text{NN}}$, the two-local nearest-neighbour Pauli set used
for the Heisenberg model in Sec.~\ref{sec:heisenberg}, included here for
comparison across model types. The same qualitative pattern recurs
throughout: cardinality is set almost entirely by whether a pool contains
the full one-body sector ($O(n^2)$, for every $\mathcal{S}_1$-based pool)
or is restricted to the Hamiltonian's own terms or a fixed locality class
($O(n)$, for $\mathcal{S}_H$ and $\mathcal{S}_{2\text{NN}}$), while the
number of QWC measurement groups is set almost entirely by \emph{locality}
rather than cardinality: every QEB or fixed-locality pool collapses to a
handful of groups independent of $n$, whereas every pool built from the
raw (JW-string) hopping operators requires $O(n)$ or $O(n^2)$ groups
regardless of how few or many operators it contains. The two properties
are therefore governed by different structural choices; which
correlations are tracked, versus how locally they are represented and
neither can be inferred from the other.
\begin{table}[t]
\centering
\small
\begin{tabular}{@{}p{2.1cm} c c@{}}
\toprule
\textbf{Pool} & \textbf{Card.} & \textbf{QWC groups} \\
\midrule
$\mathcal{S}_1$ & $O(n^2)$ & $n^2/2$ \\
$\mathcal{S}_1^{\mathrm{QEB}}$ & $O(n^2)$ & $2$ \\[3pt]
$\mathcal{S}_1{+}ZZ_{\mathrm{imp}}$ & $O(n^2)$ & $\approx n^2/2$ \\[3pt]
$\mathcal{S}_1{+}ZZ_{\mathrm{all}}$ & $O(n^2)$ & $\approx n^2/2$ \\
$\mathcal{S}_1^{\mathrm{QEB}}{+}ZZ_{\mathrm{all}}$ & $O(n^2)$ & $3$ \\[3pt]
$\mathcal{S}_H$ & $O(n)$ & $n{+}1$ \\
$\mathcal{S}_H^{\mathrm{QEB}}$ & $O(n)$ & $3$ \\[3pt]
$\mathcal{S}_{2\text{NN}}$ & $O(n)$ & $9$ \\[3pt]
$\mathcal{S}_1^{\mathrm{QEB}}{+}ZZ_{\mathrm{all}}$ + ZIZ & $O(n^2)$ & $3$ \\
\bottomrule
\end{tabular}
\caption{Hierarchy of operator pools benchmarked in Fig.~\ref{fig:12lucj}.
Cardinality is the total size of the named pool (not incremental over the
previous row); all $\mathcal{S}_1$-based pools remain $O(n^2)$ since
$\mathcal{S}_1$ dominates its own augmentations. QWC groups are counted by
a greedy qubit-wise-commuting grouping of the pool's Pauli strings
(numerically verified for $n=8$--$24$); values marked $\approx$ vary by at
most one group across the tested range, while the
$\mathcal{S}_1^{\mathrm{QEB}}{+}ZZ_{\mathrm{all}}$+ZIZ value is exact
(unchanged from $\mathcal{S}_1^{\mathrm{QEB}}{+}ZZ_{\mathrm{all}}$) since
$Z_iI_{i+1}Z_{i+2}$ is diagonal. For $\mathcal{S}_{2\text{NN}}$, cardinality is
$3n+9(n-1)=12n-9$ (single-qubit Paulis on every site plus all nine
Pauli-flavor combinations on each nearest-neighbor bond of a 1D chain);
its 9 QWC groups correspond to the 9 two-body flavors
$\{X,Y,Z\}\times\{X,Y,Z\}$, each of which threads into a single
qubit-wise-commuting chain along the lattice and absorbs the compatible
single-body terms, a count numerically confirmed stable for
$n=8$-$40$.}
\label{tab:pools}
\end{table}

\subsection{Simulator training at $20$ qubits: subspace growth with evolution time}
\label{app:20qb-training}

\begin{figure*}[t]
    \centering
     \includegraphics[width=\textwidth]{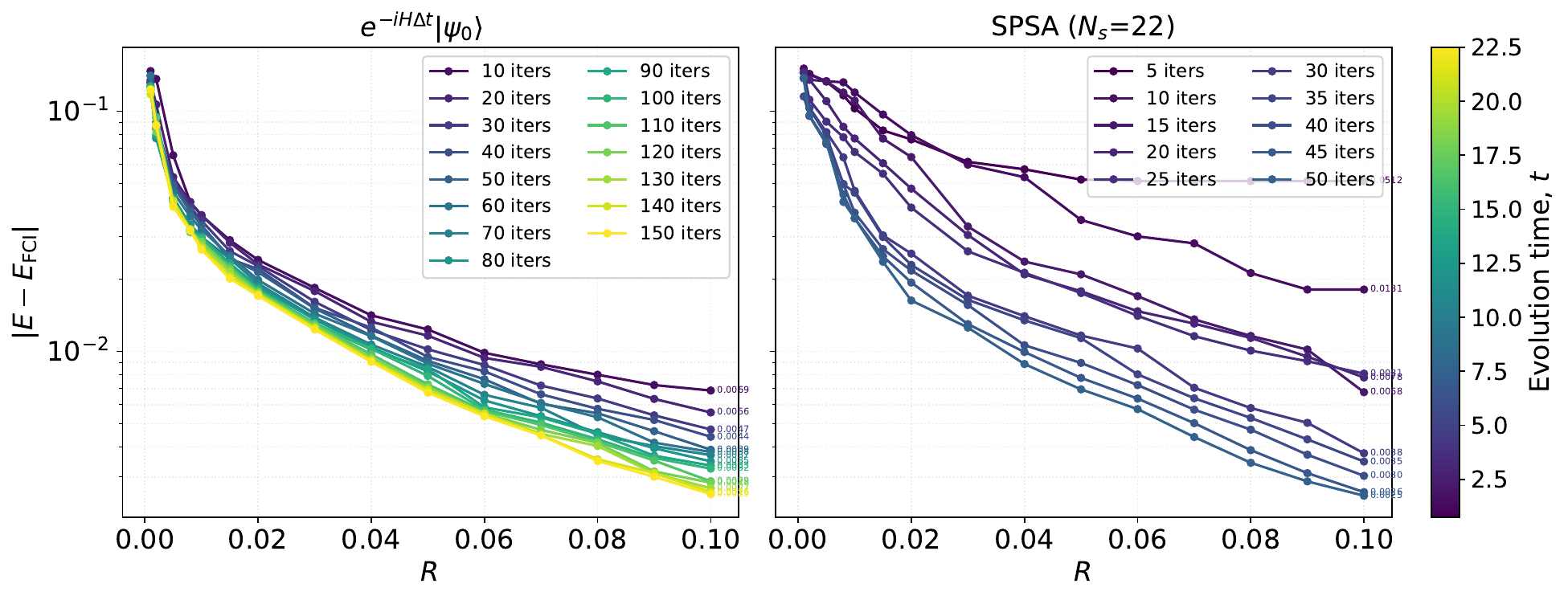}
    \caption{\textbf{QSCI subspace convergence at $20$ qubits, simulator.}
    Energy error $|E-E_{\rm FCI}|$ versus cutoff fraction of the subspace size $R$, colored by the number of
    evolution iterations. (Left) State evolved by exact matrix
    exponentiation $e^{-i\hat Ht}$, sampled and passed to QSCI at each of
    $10$--$150$ iterations. (Right) State evolved by OVQRTE with SPSA
    gradient estimation at $N_s=22$ samples per step, sampled and passed
    to QSCI at each of $5$--$50$ iterations.}
    \label{fig:20qb-training}
\end{figure*}

To isolate the effect of the evolution algorithm from that of hardware
noise, we first validate the OVQRTE-QSCI pipeline entirely on a simulator
at $20$ qubits (Fig.~\ref{fig:20qb-training}). We compare two ways of
generating the evolved state fed into QSCI at each iteration. In the
left panel, the state is propagated by direct matrix
exponentiation, $|\psi(t)\rangle=e^{-i\hat Ht}|\psi_0\rangle$, giving the
reference trajectory an ansatz-error-free upper bound on what QSCI can
achieve at a given evolution time. In the right panel, the state is
instead obtained from OVQRTE with SPSA gradient estimation at only
$N_s=22$ samples per step, which is the practical, low-cost estimator used
throughout this work. In both cases the evolved state at each iteration
is sampled into bitstring counts, which are accumulated into a growing
configuration-interaction subspace and passed to QSCI; the resulting
energy error $|E-E_{\rm FCI}|$ is plotted against the subspace size,
expressed as a percentage of the full CI space, with curves colored by
the number of iterations (i.e. the total evolution time reached).

In both panels the energy error decreases monotonically with subspace
size and with the number of iterations: later evolution times generate
configurations that are progressively more relevant to the ground state. For the exact
reference, this improvement saturates quickly, with the curves for
different iteration counts collapsing onto a narrow band once
$R/\text{full CI}\gtrsim4\%$. The SPSA-driven trajectory shows a wider
spread at low iteration counts, reflecting on the additional variance of
the stochastic gradient but converges toward the same accuracy
regime as the exact reference as the number of iterations grows. By
$50$ SPSA iterations, the energy error at $R/\text{full CI}=10\%$
reaches $\sim3\times10^{-3}$, comparable to the exact-RTE curves in the
same subspace-size range. That this level of accuracy is reached using
only $N_s=22$ samples per gradient step demonstrates that the
low-cost SPSA estimator is sufficient to drive the subspace-generation
process effectively, without requiring the exact (and far more
expensive) gradient evaluation.

\end{document}